\documentclass[twocolumn,showpacs,preprintnumbers,amsmath,amssymb,aps,prb,longbibliography,superscriptaddress]{revtex4-2}

\pdfoutput=1
\usepackage[pdftex]{graphicx}

\usepackage[english]{babel}
\usepackage{hyperref}
\usepackage{amsmath}
\usepackage{float}
\usepackage{soul}

\usepackage{amssymb}
\usepackage[utf8]{inputenc}
\usepackage[T1]{fontenc}
\usepackage{graphicx}
\usepackage{braket}

\usepackage[caption=false]{subfig}

\usepackage{cancel}
\usepackage{color,soul}

\usepackage{epstopdf}
\usepackage{enumitem,kantlipsum}
\usepackage{physics}
\usepackage{bbold}
\usepackage{array}

\usepackage[T1]{fontenc}
\usepackage[sc]{mathpazo}
\makeatletter

\usepackage[caption=false]{subfig}
\usepackage{latexsym}
\usepackage{braket}

\@ifundefined{showcaptionsetup}{}{%
 \PassOptionsToPackage{caption=false}{subfig}}
\usepackage{subfig}
\makeatother

\begin{document}
\title{Long-range hopping and indexing assumption in one-dimensional topological insulators}
\author{R. G. Dias}
\email{rdias@ua.pt}

\affiliation{Department of Physics $\&$ i3N, University of Aveiro, 3810-193 Aveiro,
Portugal}
\author{A. M. Marques}
\affiliation{Department of Physics $\&$ i3N, University of Aveiro, 3810-193 Aveiro,
Portugal}
\date{\today}
\begin{abstract}
In this paper, we show that the introduction of long-range hoppings
in  1D topological insulator models implies that different
possibilities of site indexing must be considered 
when determining the bulk topological invariants in order to avoid the existence of hidden symmetries.
The particular case of the extended SSH chain is addressed as an example where such behavior occurs. In this model, the introduction of long-range hopping terms breaks the bipartite property  and a band inversion occurs in the band structure as the relative values of the hopping terms change, signaling a crossover between hopping parameter regions of ``influence'' of different chiral symmetries.
Furthermore,   edge states  become a linear combination of  edge-like states with different localization lengths and reflect
the gradual transition between these  different chiral symmetries.

\end{abstract}

\maketitle
\section{Introduction}

Topological insulators (and superconductors) are categorized according to their dimension and to general symmetries that protect gapless boundary modes \cite{Altland1997,Chiu2016}. One of the latter is the chiral symmetry or sublattice symmetry which protects
the zero energy edge states of  bipartite lattices such as the SSH model  due to the anticommutation of the chiral operator ${\cal C} $ (the difference of the projection
operators in the two sublattices of the system) and the Hamiltonian.
It is usual in the case of translational invariant Hamiltonians to assume that the latter is equivalent to the relation ${\cal C} H(k) {\cal C}^{-1}=- H(k)$, where implicitly a choice of unit cell has been undertaken.
Furthermore, several topological invariants such as the Zak's
phase \cite{Zak1989} or the Chern number \cite{Asboth2016}  rely on the knowledge of  Bloch eigenstate $\ket{u_{\pm}(k)}$ throughout the Brillouin zone and also  reflect a particular choice of unit cell. Each choice of unit cell implies a particular subdivision of the lattice into sublattices.

In this paper, we  argue that in the case of linear chains with long-range hoppings (beyond nearest neighbor and that may break  the bipartite property),  several  choices of site  indexing should be considered in order to correctly describe their bulk topological behavior, which correspond to \emph{different representations of the tight-binding Hamiltonian  as a linear chain}.
If the indexing is not the ``correct''  one, the topological protection of edge states may be associated  with a \emph{hidden chiral symmetry} \cite{Li2015,Zurita2021,Fukui2013,Hou2016,Hou2017}, which cannot be written as a difference of  projection
operators onto the sublattices defined by the unit cell. This will be expected in particular if a band inversion occurs in the band structure of the linear chain  as the relative values of the hopping terms change.
In order to illustrate the previous arguments, we consider in this paper the chiral symmetry protected
topological SSH insulator and add to it long-range
 hopping terms \cite{Hsu2020,Maffei2018,Chen2020,Ahmadi2020,Zhang2017,Li2019,Xie2019,Fu2020,PerezGonzalez2019,PerezGonzalez2018,Longhi2018,Hetenyi2021,Li2014} in such a way that the Hamiltonian eigenbasis remains
the same  and the usual SSH chiral operator remains a mapping operator between eigenstates of the Hamiltonian (allowing us to compare it with new chiral operators).  This model is simply the extended SSH chain with next-nearest neighbor (NNN) hoppings. 
For certain evolutions of the ratio between hopping parameters, a band inversion will occur signaling a crossover between parameter regions of ``influence'' of different chiral symmetries (regions where the Hamiltonian can be adiabatically changed in order to recover  a chiral symmetry).

We  also present an analytical and numerical study of the topological edge states of this  extended SSH chain in the presence of open boundary conditions (OBC), and show that they reflect clearly the indexing problem and the existence of different chiral symmetries. 
The extended hopping terms imply that  the edge states can not be generated from degenerate bulk states and this leads to the existence of two localization lengths in the real-space dependence of the edge states. In a Log-plot, such behavior is clearly observed as well as the existence of  competition between the different chiral symmetries in the edge state profile in the crossover region. 

This paper is organized in the following way.
In Section II, we  introduce in general terms the indexing problem. In the next Section, we construct the extended SSH model and discuss the indexing problem in this model. In Section IV, we address limiting cases  of the extended SSH model where a chiral symmetry is present. In Section IV, we present a discussion of the edge states in this model. Finally we conclude.

\begin{figure}[t]
\includegraphics[width=.45\textwidth]{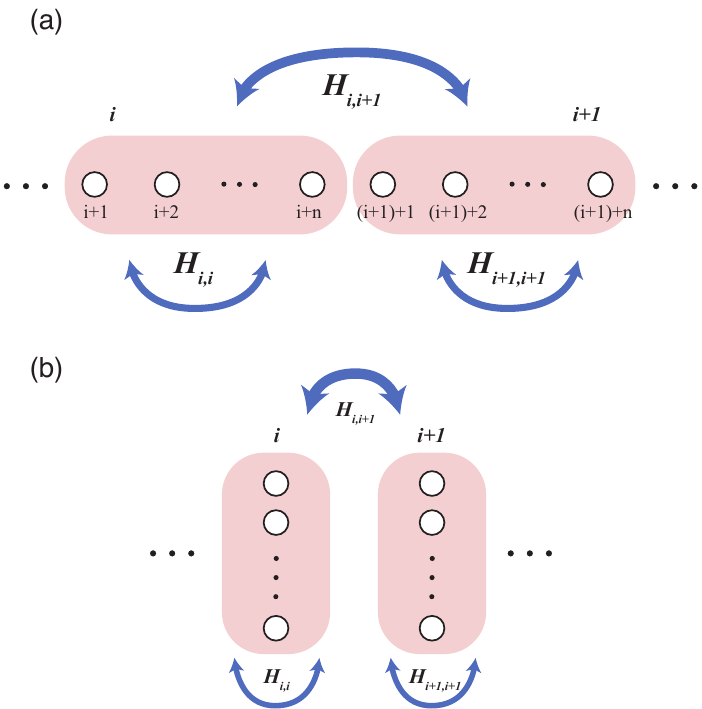}\caption{ (a) Linear chain with finite range extended hopping terms and a unit cell with $n$ sites such that only  hopping terms within the unit cells ($H_{i,i}$) or between consecutive unit cells ($H_{i,i+1}$) are present. (b) Reinterpretation  of the linear chain as a ladder with a rung with $n$ sites. Different indexings of the ladder sites generate different linear chains. 	\label{fig:ladder}}
\end{figure}

\begin{figure}
	\begin{centering}
		\subfloat[\label{fig:index1}]{\begin{centering}
				\includegraphics[width=0.9\columnwidth]{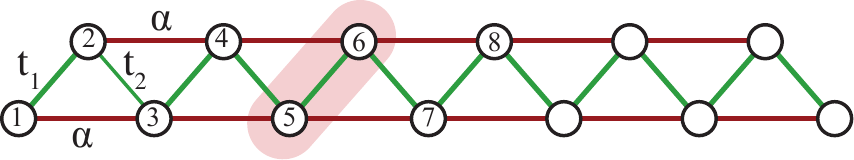}
				\par\end{centering}
		}
		\par\end{centering}
	\begin{centering}
		\subfloat[]{\begin{centering}
				\includegraphics[width=0.9\columnwidth]{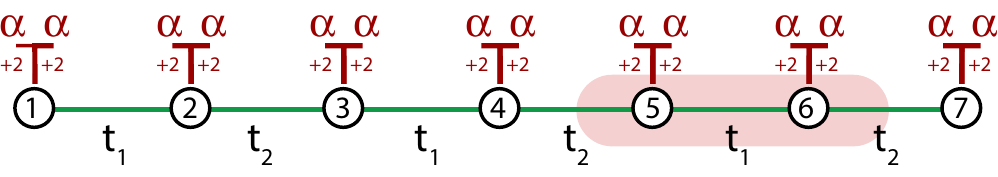}
				\par\end{centering}
		}
		\par\end{centering}
	\begin{centering}
		\subfloat[\label{fig:index2}]{\begin{centering}
				\includegraphics[width=0.9\columnwidth]{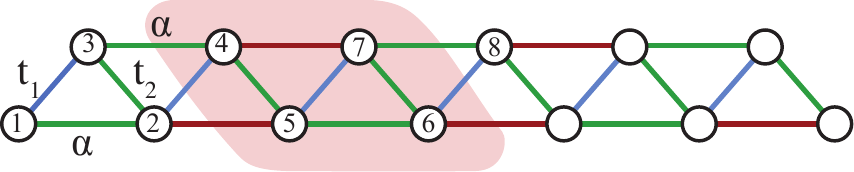}
				\par\end{centering}
		}
		\par\end{centering}
	\centering{}\subfloat[]{\begin{centering}
			\includegraphics[width=0.9\columnwidth]{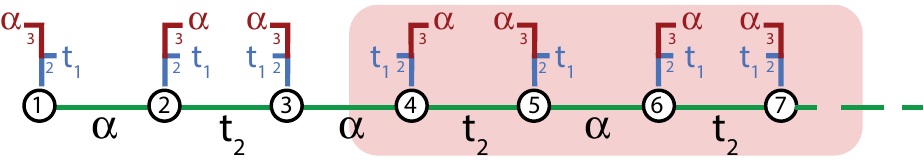}
			\par\end{centering}
	}\caption{(a) Usual indexing of the SSH chain with next-nearest-neighbor hopping $\alpha$.
		(c) New indexing with the indexing being flipped every other rung. 
		Note that for finite $t_{1}$ the unit cell has four sites (and we should consider a ladder with a 4-site rung). 
		(b) and  (d): Linear chain
		corresponding to the indexing in (a) and (c), respectively, using a shortened representation for the  long-range hoppings, whose numbers indicate the distance between connected sites. In (b) [(d)], the red (blue) hopping terms break the
		bipartite condition (that is, they are hopping terms between sites on the same sublattice).\label{fig:Usual-indexing-1}}
\end{figure}

\begin{figure*}[t]
	\centering{}\includegraphics[width=.8\textwidth]{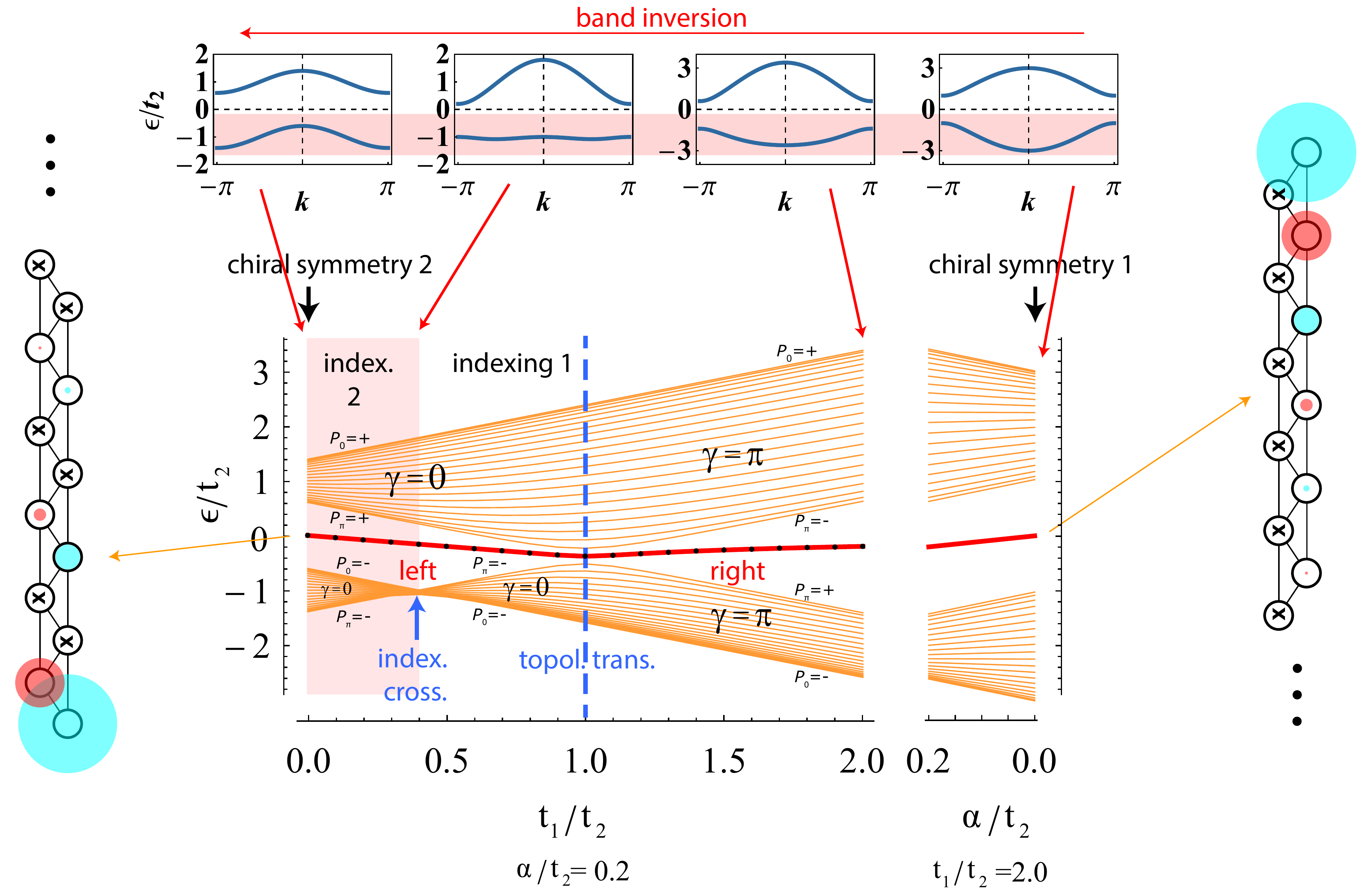}\caption{ On the left, spectrum as a function of $t_1/t_2$ (with $\alpha/t_2=0.2$)   for the SSH tight-binding model with next-nearest neighbor hopping $\alpha$ for a chain with 21 unit cells plus one site [with a $t_1$ ($t_2$) link at the left (right) edge] and adopting the indexing of Fig.~\ref{fig:Usual-indexing-1}(a). On the right,  spectrum as a function of $\alpha/t_2$ (with $t_1/t_2=2$).
	The left and right limiting cases show a chiral spectrum, but with different chiral symmetry reflecting a different unit cell or equivalently, different choices of sublattices.  The need for a different indexing in these two cases is signaled by the bottom  band   inversion and is evident in the amplitude profile of the corresponding edge states shown on the left (for visual clarity, we show instead the edge state  corresponding to $\alpha/t_2=0.4$) and on the right. The circle sizes indicate the relative amplitudes and the colors reflect the phases (blue is zero, red is $\pi$). 	Parity values and the respective Zak's phases are shown assuming the indexing of   Fig.~\ref{fig:Usual-indexing-1}(a).
		\label{fig:sshnnn}}
\end{figure*}

\section{The indexing assumption}
Let us consider a linear chain with finite range extended hopping terms and a unit cell with $n$ sites, where $n$ is even. Instead of the shortest unit cell associated with the translational symmetry, we consider a larger unit cell  with size at least large enough so  that  hopping terms within the unit cells and between consecutive unit cells are present, see Fig.~\ref{fig:ladder}(a). This chain can be interpreted as a ladder with a rung of $n$ sites as shown in  Fig.~\ref{fig:ladder}(b). Our chain has $n$ sublattices (all with same cardinality) and the site $j$ belongs to the sublattice  $[(j+1) \text{mod }n] -1$. If  the ladder was our starting point, then the original linear chain would be obtained with a zig-zag indexing of the sites in the ladder  shown in Fig.~\ref{fig:ladder}(b), that is, our linear chain constitutes a particular path in our ladder that generates a periodic sequence of nearest-neighbor (NN) hoppings as well as of the longer range hopping terms. The respective periodicity defines the unit cell of the linear chain and it should be obvious that an infinite number of different paths exist, but in general with larger unit cells. 

Starting from a ladder and given some hopping terms distribution, it should be possible to find the path (or paths) corresponding to the shortest unit cell (which may be smaller than the number of sites in each rung). Given one of these paths, one can circularly permutate the indexing of sites in a given sublattice without increasing the unit cell, but at the cost of introducing longer range hopping terms in general. If we permutate in a non-circular way the sites in a given sublattice, the unit cell size will increase or one may even lose completely the translational invariance in the linear chain (but a hidden translational invariance will be present in the Hamiltonian). 
One would naturally assume that the  indexing corresponding to the shortest unit cell would  allow for the full topological characterization of the linear chain, but there is a subtle problem concerning the protecting symmetries of the edge states when our linear chain has long-range hopping that we explain in the following. 

Given our ladder of $n$-sites per rung, it is possible to change the hopping parameters so that a single path is present such that all sites are only  connected to their nearest neighbors in the corresponding linear chain. This chain is bipartite (with the odd and even sublattices having the same cardinality) and therefore it has chiral symmetry reflecting the existence of hopping terms only between even and odd sublattices. 
One may ask how many different such chiral symmetries (as a simplification, we consider only the  chiral symmetries reflecting a partition of the lattice into two sublattices with the same cardinality, but the argument can in principle be generalized to any ratio of the sublattice cardinalities)  can be found given an $n$-site unit cell (the ladder rung) and allowing only changes in the magnitudes of the hopping terms (no on-site potentials are introduced). This can be answered considering the possible linear paths (with only nearest neighbor hopping) that generate different sublattices. Note that these linear paths include a single hopping term between consecutive unit cells since we can do a circular permutation of indexing within the sublattices in order to ensure that. The number of different chiral symmetries is $n(n-1)/2$ and the simplest topological linear chain corresponding to each chiral symmetry is an SSH chain with staggered hopping terms. 

Given a particular linear path and the respective NN linear chain, the introduction of long-range hopping terms between the odd and even sublattices will not break the chiral symmetry of the linear chain, so one can say  each chiral symmetry corresponds to a particular region of the ladder  tight-binding parameter space. Furthermore besides the region where the chiral symmetry is present, there is a region of ``influence'' of chiral symmetry  which is the region in the hopping parameter space  of the general Hamiltonian (that does not have chiral symmetry due to the long-range hoppings) where the Hamiltonian can be adiabatically deformed into the Hamiltonian with chiral symmetry.

In the general case of the ladder, one will have several coexisting choices of linear paths with larger and smaller unit cells as we will see in the next section. If we start from a ladder Hamiltonian with  chiral symmetry and the respective  shortest unit cell is considered, as one varies the hopping terms, one may approach another chiral symmetry point in the parameter space. If one wants to avoid the existence of hidden symmetries in the respective topological characterization, there are two possibilities: (i) to consider a larger unit cell such the new chiral operator can be represented in a single unit cell; (ii) to change the indexing of sites in order to obtain the  unit cell corresponding to the chiral symmetry point. 
In the next section, we will illustrate all these arguments in the case of the extended SSH chain.

\section{Model}
In this section, we arrive to the NNN SSH model following a particular reasoning motivated by the indexing problem, that is, we start from the premise that the introduction of long-range hopping should not change the Hamiltonian eigenbasis (it should remain 
the same as that of the usual SSH chain) in order for the usual chiral operator of the SSH chain to remain a mapping operator between the same eigenstates of the Hamiltonian. This will allow for a comparison between that operator with the new one  associated with a new chiral symmetry, see Section IV.
Note that the extended SSH has been addressed in the past in several works \cite{Hsu2020,Maffei2018,Chen2020,Ahmadi2020,Zhang2017,Li2019,Xie2019,Fu2020,PerezGonzalez2019,PerezGonzalez2018,Longhi2018,Hetenyi2021,Li2014}, but as far as we know its indexing problem has never been discussed.

Let us then consider a 1D tight-binding model with
non-symmetric bands $\varepsilon_{k}^{\pm}$, but with the same eigenbasis
of the SSH model, $\{\ket{u_{\pm}(k)}\}$. Note that since we fixed the
eigenbasis, the usual SSH chiral operator $\hat{C}_{1}\vert\psi_{k}^{\pm}\rangle=\vert k\rangle\otimes\hat{C}_{k}\ket{u_{\pm}(k)}$
is still valid as a mapping operator between eigenstates (a given state in one band has
the respective pair in the other band). This chiral operator is written  in terms of the  Bloch eigenstates $\ket{u_{\pm}(k)}$ of the Hamiltonian
$H_{k}$ as  $\hat{C}_{1}=\sum_{k}\hat{C}_{1}(k)$ with $\hat{C}_{1}(k)=\ket{u_{+}(k)}\bra{u_{-}(k)}+H.c.$,
with $H_{k}\ket{u_{\pm}(k)}=\varepsilon_{\pm}(k)\ket{u_{\pm}(k)}$, and $\varepsilon_{+}(k)=-\varepsilon_{-}(k)$ if the bands are symmetric. 
If the bands are non-symmetric, one loses
the anticommutation relation between
this  operator and the Hamiltonian (which is no longer bipartite). This implies that the  mapping operator
 is still given by $\hat{C}_{1}(k)=\ket{u_{+}(k)}\bra{u_{-}(k)}+H.c.$, but with $\varepsilon_{+}(k)\neq-\varepsilon_{-}(k)$.
Furthermore, topological invariants such as the Zak's phase that are calculated from this eigenbasis remain exactly the same with quantified values and a discontinuity when  $t_1=t_2$. The matrix representation
of $H(k)$ in this eigenbasis is $H(k)=diag(\varepsilon_{+}(k),\varepsilon_{-}(k))$
where $\varepsilon_{+}(k)>\varepsilon_{-}(k),\forall k,$
except at particular choice of the parameters where the gap closes.
The matrix representation in the ``Wannier'' basis $\{\vert k;A\rangle,\vert k;B\rangle\}$
is 
\begin{eqnarray}
H(k) & = & \begin{bmatrix}\varepsilon_{d}(k) & \varepsilon_{o}(k)e^{-i\phi(k)}\\
\varepsilon_{o}(k)e^{i\phi(k)} & \varepsilon_{d}(k)
\end{bmatrix},
\label{eq:hamilt}
\end{eqnarray}
where we have diagonal terms $\varepsilon_{d}(k)=[\varepsilon_{+}(k)+\varepsilon_{-}(k)]/2$
and off-diagonal terms $\varepsilon_{o}(k)=[\varepsilon_{+}(k)-\varepsilon_{-}(k)]/2$.
The above matrix has no $\sigma_z$ component and that implies that a winding vector can still be defined in the $xy$ plane.

In the case of the usual SSH model, we have $\varepsilon_{o}(k)e^{i\phi(k)}=t_{1}+t_{2}e^{iak}$
and $\varepsilon_{d}(k)=0,$ and therefore $\varepsilon_{\pm}(k)e^{i\phi(k)}=\pm(t_{1}+t_{2}e^{iak}).$
The Fourier transform of the diagonal terms is zero and the off-diagonal
term is $\frac{1}{N}\sum_{k}e^{ik(j-l)}(t_{1}+t_{2}e^{iak})=t_{1}\delta_{jl}+t_{2}\delta_{j,l+1}$
and $\frac{1}{N}\sum_{k}e^{ik(j-l)}(t_{1}+t_{2}e^{-iak})=t_{1}\delta_{jl}+t_{2}\delta_{j,l-1}$,
so we obtain correctly the SSH model.

In the general case of the Hamiltonian in Eq.~\ref{eq:hamilt}, we will have long-range hoppings
and onsite potentials. In this paper, we consider a simple case where
we keep the SSH model conditions except for $\varepsilon_{+}(k)+\varepsilon_{-}(k)=4\alpha\cos(ka)$, that is,
\begin{eqnarray}
H(k) & = & \begin{bmatrix}
2\alpha\cos(ka)& t_{1}+t_{2}e^{-iak}\\
t_{1}+t_{2}e^{iak} & 2\alpha\cos(ka)
\end{bmatrix}.
\end{eqnarray}
The real-space Hamiltonian in the Wannier basis $\{\vert j;A\rangle,\vert l;B\rangle\}$
is obtained from $H_{jl}=\frac{1}{N}\sum_{k}e^{ik(j-l)}H(k)$. Since
$\frac{1}{N}\sum_{k}e^{ik(j-l)}(\varepsilon_{+}(k)+\varepsilon_{-}(k))=\alpha(\delta_{j,l+1}+\delta_{j,l-1})$
and
\begin{eqnarray}
\varepsilon_{o}(k)e^{i\phi(k)} & = & t_{1}+t_{2}e^{iak}\\
 & = & \sqrt{2t_{1}t_{2}\cos(ak)+t_{1}^{2}+t_{2}^{2}}e^{i\phi(k)},\nonumber 
\end{eqnarray}
one obtains an SSH tight-binding model with next-nearest neighbor
hoppings, as depicted in Fig.~\ref{fig:Usual-indexing-1}(a), whose real-space Hamiltonian is written as
\begin{eqnarray}
H=\sum_{j} &[&t_{1}c_{jA}^{\dagger}c_{jB}+t_{2}c_{jB}^{\dagger}c_{j+1A}\\
 &+&\alpha(c_{jA}^{\dagger}c_{j+1A}+c_{jB}^{\dagger}c_{j+1B})+H.c.].
\end{eqnarray}
The respective bands are $\varepsilon_{\pm}(k)=\varepsilon_{d}(k)\pm\left|\varepsilon_{o}(k)\right|$. 

For general values of the hopping parameters, this model falls into the AI symmetry class due the absence of a chiral symmetry. 
The chiral symmetry is recovered when the chain becomes bipartite, moving this model into the BDI class \cite{Altland1997,Chiu2016}, that is, it has time reversal, particle-hole, and chiral symmetries.

\subsection{Spectrum of the extended SSH chain}
The introduction of  NNN hoppings destroys the bipartite property of the SSH chain
and this is reflected by the fact that the spectrum is no longer symmetric.
In Fig.~\ref{fig:sshnnn}, we show the evolution of the spectrum  for the SSH tight-binding model with NNN hopping $\alpha$ for a chain with 21 unit cells plus one site and adopting the indexing of Fig.~\ref{fig:Usual-indexing-1}(a). On the right side,  $\alpha=0$ and one has the usual SSH chain with hopping parameters $t_1/t_2=2$. This chain has a right edge state (shown on the right of Fig.~\ref{fig:sshnnn}), where the the usual subdivision into two sublattices is evident from the fact the edge state has zero amplitudes on one sublattice (indicated by the x-symbols), reflecting the usual SSH chiral symmetry. This edge state survives as we increase $\alpha$, but gains finite amplitude in the latter sublattice. In Fig.~\ref{fig:sshnnn}, we increase  $\alpha$ up to 0.2 and then, keeping $\alpha$ fixed, we change $t_1$ from  $t_1/t_2=2$ to  $t_1/t_2=0$ (left limit).  The left  limiting case again shows a chiral spectrum but with different chiral symmetry reflecting a different unit cell or, equivalently, different choices of sublattices.  The need for a different indexing in the left limiting case  is signaled by the bottom  band   inversion at $t_1/t_2\approx 0.4$ and is evident in the amplitude profile of the corresponding edge state shown on the left (for visual clarity, we show instead the edge state  corresponding to $\alpha/t_2=0.4$).	
This band inversion implies multiple level crossings (as shown in Fig.~\ref{fig:sshnnn}) and this implies that the usual reasoning of adiabatic transformation  can only be applied from the left (right) to this indexing crossover point. That is, the topological properties to the left of this point ($t_1/t_2<0.4$) in Fig.~\ref{fig:sshnnn} cannot be adiabatically connected to those of the SSH model (with $\alpha=0$).

Parity values and the respective Zak's phases are also shown in Fig.~\ref{fig:sshnnn} assuming the indexing of Fig.~\ref{fig:Usual-indexing-1}(a). Since the Bloch eigenbasis remains the same throughout this evolution, they are the same of the usual SSH chain, except for the change in parity associated with the bottom band inversion. As in the case of the usual SSH model, a discontinuity in the
phases of the components of $\ket{u_{1}(k)}$ occurs (assuming a smooth gauge) when
the gap is closed at $t_1=t_2$.
At this point, the Zak's phase is also discontinuous (and quantified),
since it is calculated from the eigenstates of the Hamiltonian (or
equivalently, the parity of the $k=\pi$ states). Another discontinuity
is present at the phase of the off-diagonal matrix element of the
Block Hamiltonian. Note that since the eigenstates are the same as
that of the usual SSH model, the Hamiltonian has inversion symmetry.

The bottom band inversion is also evident in the left top plot of Fig.~\ref{fig:sshnnn} as well as the need for a $\pi$ shift in the bottom band so that the spectrum becomes symmetric as expected due to the presence of a chiral symmetry.
It is not possible to describe the latter chiral symmetry using the unit cell of  Fig.~\ref{fig:Usual-indexing-1}(a), even despite the fact that this remains a perfectly valid unit cell when NNN hoppings are introduced. Therefore, in the general case where $\alpha\neq0$, other indexings
may need to be considered depending on the relative values of the hopping terms. In Fig.~\ref{fig:Usual-indexing-1}(c), we
show a different possible indexing that implies a 4-site unit cell.
In Figs.~\ref{fig:Usual-indexing-1}(b) and (d), the indexings of (a) and (c) are represented in a linear chain where we indicate the long range hoppings using a shortened representation (that may be clearer  if one wishes to increase their number).

\section{Indexing and chiral symmetry in limiting cases}
To recap, there are two simple limits of our extended SSH model where a chiral symmetry is present:

\subsection{The nearest-neighbor SSH chain ($\alpha\mathbf{=0}$)}
Setting $\alpha=0$ in our extended SSH chain of Fig.~\ref{fig:Usual-indexing-1}, there is a natural indexing of sites which is displayed in  Fig.~\ref{fig:Usual-indexing-1}(a). The
maximum number of finite Hamiltonian matrix elements in each row (this is equivalent the coordination number of each site) is two, then one
has a NN tight-binding model and one numbers the sites following the path of finite matrix elements. For this indexing, this bipartite  model has a sublattice of even sites connected to the odd sublattice sites in such a way that $t_{ij}\neq0$ if $\vert i-j\vert=1$. 
This model has a chiral symmetry and the respective chiral operator is the difference of the projection
operators into the even and odd sublattices.
This description can be applied to any other limit of the model where a single path of hopping terms is present.

\subsection{The other chiral case $\mathbf{t_{1}=0}$}
Assuming an even stricter condition,  $t_{1}=\alpha=0$,
one has decoupled dimers and  two flat bands of energies $t_{2}$
and $-t_{2}$ corresponding respectively to bonding and anti-bonding
dimer states. These dimers can be connected in several ways in order
to generate chains with only NN hoppings. One is to
add horizontal hoppings alternating between the top and the bottom [green lines in  Fig.~\ref{fig:Usual-indexing-1}(c)]
and another is to do it as in the SSH model [green lines in  Fig.~\ref{fig:Usual-indexing-1}(a)]. This leads to the different
indexings in Figs.~\ref{fig:Usual-indexing-1}(c) and (a). These two ways
obviously imply different chiral operators, reflecting the choice
of even and odd sublattices.

Considering $t_{1}=0$, $\alpha\neq0$, and the indexing of Fig.~\ref{fig:Usual-indexing-1}(a), the Hamiltonian is $H(k)=2\alpha\cos(ak)\hat{1}+t_{2}\hat{\sigma}_{x}$, where $t_2$ is assumed as an intracell coupling, and
the system can be mapped onto two simple tight-binding chains {[}of
bonding (anti-bonding) states with onsite potentials $t_{2}$ ($-t_{2}$)
and hopping constant $\alpha$]. These two chains have identical band dispersions
apart from the energy shift $2t_{2}$,
\begin{eqnarray}
\varepsilon_{\pm}(k)/\vert t_{2}\vert & = & \frac{\alpha}{\vert t_{2}\vert}\cos(ka)\pm1,
\end{eqnarray}
with $\ket{u_{\pm}(k)}=\begin{bmatrix}\pm1 & 1\end{bmatrix}^{T}/\sqrt{2}$.
Apparently this model does not have a chiral symmetry and the absence
of a $k$-dependent phase in $\ket{u_{\pm}(k)}$   may lead one to
assume incorrectly that no topological phase can be present. Despite the
fact that at the gap closing point, $t_{1}=0$ and $\alpha/\vert t_{2}\vert=.5$,
the bands do not touch, a topological transition does occur at this
point with the appearance of edge states (note that if one considers
a four-site unit cell, this leads to the folding of the bands and the
bands touch at the gap closing point).
Again, this reflects the fact that
the model is  bipartite in this limit, but the respective sublattices
are not the same as those of the initial SSH model and the respective
translational operator of the two-site unit cell is also not the same
(see Fig. \ref{fig:Usual-indexing-1}). Therefore in order to correctly understand
the topological phase, the new indexing of the sites of Fig. \ref{fig:Usual-indexing-1}(c)
is required.

Let us denote the indexing of Fig. \ref{fig:Usual-indexing-1}(a)
by $n$ and the indexing of the sites of Fig. \ref{fig:Usual-indexing-1}(c)
by $\bar{n}$. Considering the 4-site unit cell of Fig. \ref{fig:Usual-indexing-1}(c) [note that for $t_1=0$,  one has a 2-site unit cell  with the indexing of Fig. \ref{fig:Usual-indexing-1}(c) as easily concluded from Fig. \ref{fig:Usual-indexing-1}(d), but the 4-site unit cell makes the change of indexing easier to understand],
then the bonding ($+$) and anti-bonding ($-$) states are
\begin{equation}
\vert\psi_{n}(k)\rangle_\pm=\vert2k\rangle\otimes\begin{bmatrix}\psi_{4}\\
\psi_{5}\\
\psi_{6}\\
\psi_{7}
\end{bmatrix}_\pm=\frac{1}{2}\vert2k\rangle\otimes\begin{bmatrix}\pm1\\
1\\
\pm e^{ika}\\
e^{ika}
\end{bmatrix},
\end{equation}
and using the new indexing they become
\begin{eqnarray}
\vert\psi_{\bar{n}}(k)\rangle_\pm&=&\vert2k\rangle\otimes\begin{bmatrix}
\psi_{\bar{4}}\\
\psi_{\bar{5}}\\
\psi_{\bar{6}}\\
\psi_{\bar{7}}
\end{bmatrix}_\pm
=\vert2k\rangle\otimes\begin{bmatrix}\psi_{4}\\
\psi_{5}\\
\psi_{7}\\
\psi_{6}
\end{bmatrix}_\pm \nonumber
\\
&=&\frac{1}{2}\vert2k\rangle\otimes\begin{bmatrix}\pm1\\
1\\
e^{ika}\\
\pm e^{ika}
\end{bmatrix}.
\end{eqnarray}
Therefore the band of bonding states is the same for both indexings,
but in the case of the anti-bonding band the Bloch eigenstates, when using the indexing of Fig.~\ref{fig:Usual-indexing-1}(c) and a 2-site unit cell, become $\vert k+\pi\rangle\otimes\vert u_{-}(k+\pi)\rangle$
and this implies a $\pi-$shift in the momentum of the bottom band.

The bulk Hamiltonian of the chain in Fig. \ref{fig:Usual-indexing-1}(d),
assuming a unit cell of sites $\{2,3\}$ and $t_1=0$, is $H(k)=\left[t_{2}+2\alpha\cos(ak)\right]\hat{\sigma}_{x}$,
the eigenvalues are $\pm\left[t_{2}+2\alpha\cos(ak)\right]$ and the
respective eigenstates are again $\ket{u_{\pm}(k)}=\begin{bmatrix}\pm1 & 1\end{bmatrix}^{T}/\sqrt{2}$,
that is, they are independent of $k$ for this choice of unit cell.
The respective Zak's phase is zero for both bands and this is expected
since if the site 1 is absent in Fig.~\ref{fig:Usual-indexing-1}(d),
no edge state will be present. The same occurs in the case of the
old indexing of Fig. \ref{fig:Usual-indexing-1}(b), but note
that a shift of one site in the unit cell leads to a finite Zak's
phase and this agrees with the fact that for the choice of OBC of
Fig.~\ref{fig:Usual-indexing-1} an edge state is observed.

The correct chiral operator ${\cal C}_{2}$ that protects the edge
states in this case requires two two-site unit cells if one works
with the indexing of sites in Fig.~\ref{fig:Usual-indexing-1}(a),
but a single two-site unit cell using the indexing of sites in Fig.~\ref{fig:Usual-indexing-1}(c).
Using the former indexing, we have in the  Wannier basis of two
unit cells
\begin{equation}
{\cal C}_{2}=
\begin{bmatrix}1 & 0 & 0 & 0\\
0 & -1 & 0 & 0\\
0 & 0 & -1 & 0\\
0 & 0 & 0 & 1
\end{bmatrix},
\end{equation}
or in the two-band Bloch basis,
\begin{equation}
{\cal C}_{2}(k)=\vert\varepsilon_{k,+}\rangle\langle\varepsilon_{k+\pi,-}\vert+\vert\varepsilon_{k,-}\rangle\langle\varepsilon_{k+\pi,+}\vert+H.c.,
\end{equation}
or, equivalently,
\begin{equation}
{\cal C}_2 H(k) {\cal C}_2^{-1}=- H(k+\pi).
\end{equation}
Similar operators have been found in Refs.~\cite{Li2015,Zurita2021}.
This expression implies that by using the indexing of Fig.~\ref{fig:Usual-indexing-1}(a)
the chiral operator ${\cal C}_2$ cannot be defined as an operator acting only
in $\ket{u_{\pm}(k)}$ as in the case of the SSH model (unless we use
a larger unit cell with four sites) and in fact, this operator reflects a hidden chiral symmetry for that indexing. If the indexing of Fig.~\ref{fig:Usual-indexing-1}(c)
is used, the chiral operator ${\cal C}_2$ can now be written as an operator (with the usual form) acting
only in $\ket{u_{\pm}(k)}$, but not the chiral operator ${\cal C}_1$, that is, the chiral symmetry of the SSH chain becomes a hidden chiral symmetry if we choose the indexing of Fig.~\ref{fig:Usual-indexing-1}(c). Note
that this implies that, since the edge states for $t_1=0$ are eigenstates of this
chiral operator ${\cal C}_2$, the ``unit cell'' for the edge state shown at the left of Fig.~\ref{fig:sshnnn} has four sites using
the indexing of Fig.~\ref{fig:Usual-indexing-1}(a) and two sites
using the indexing of Fig.~\ref{fig:Usual-indexing-1}(c).  

\section{Edge states of the extended SSH chain}
Let
us now discuss the characteristics of the edge states in the general
case where $t_{1}\neq0$ and $\alpha\neq0$. In several recent works \cite{Delplace2011,Banchi2013,Huegel2014,Duncan2018,Marques2019,Marques2020,Fukui2020},
it has been assumed that edge states in 1D topological insulators can
be determined from the bulk Hamiltonian using the substitution $e^{ik}\rightarrow c$
in the bulk Hamiltonian, reflecting the assumption of an ansatz
\begin{equation}
\ket{\psi_{edge}}_{j}=c^{j}\ket{u(c)}=c^{j}\begin{bmatrix}\psi_{A}\\
\psi_{B}
\end{bmatrix}.
\end{equation}
(where $j$ is the unit cell index) in the eigenvalue equation of
the infinite system. Here we show that the presence of long-range
hopping in the SSH chain modifies this ansatz, due to the competition of the two
chiral symmetries associated with the two limits $t_{1}=0$ and $\alpha=0$
(reflecting the two indexings of Fig.~\ref{fig:Usual-indexing-1}
and the respective even and odd sublattices).

\begin{figure}[t]
\includegraphics[width=0.7\columnwidth]{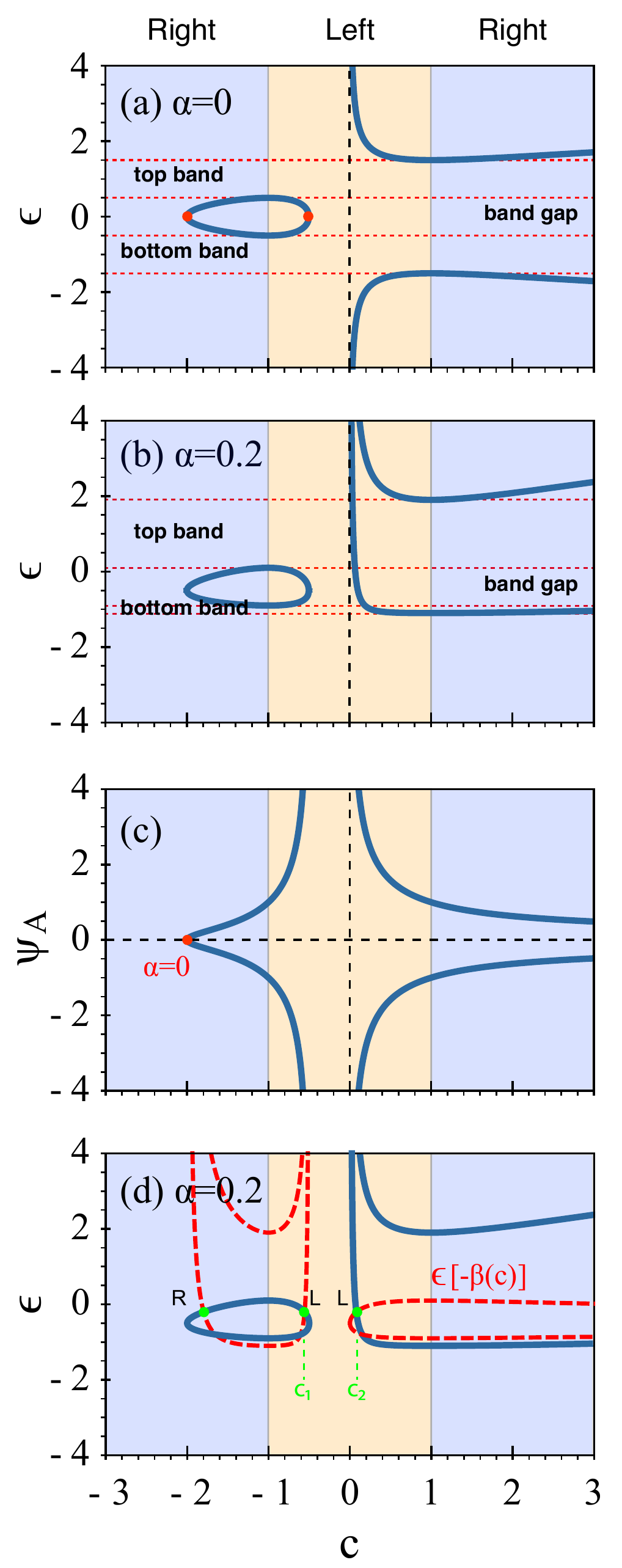}
\caption{\label{fig:cvalues}(a) and (b): real $c$ roots  (corresponding
to edge-like states of the infinite chain) of the characteristic polynomial of the $c$-matrix in Eq.~\ref{eq:cmatrix} and the respective energy
for $t_{1}=0.5$, $t_{2}=1$, (a) $\alpha=0$ or (b) $\alpha=0.2$. In the case of  $\alpha=0$, there
two possible values of $c$ for each energy in the band gap (corresponding to a right
and a left edge-like state). When NNN hopping is present ($\alpha=0.2$),
four $c$ values are present in the band gap region, with one of the right edge solutions outside the range of this plot. (c) Amplitude $\psi_A$
of the eigenstate in Eq. \ref{eq:camplitudes} as a function of real
$c$. (d) The intersection of the curves $\varepsilon_{\pm}(c)$ and $\varepsilon_{\pm}(-\gamma(c))$ gives us the $c_1$ and $c_2$ values present in the edge state [with amplitudes of the form $\psi_{A/B}(c_1^j-c_2^j)$].}
\end{figure}

Assuming the unit cell $(A=5,B=6)$ shown in Fig.~\ref{fig:Usual-indexing-1}(a),
the eigenvalue equation for the edge-like states in the infinite chain
generates the following equations
\begin{eqnarray}
\varepsilon c^{j}\psi_{A} & = & \alpha\psi_{A}c^{j-1}+\alpha\psi_{A}c^{j+1}+t_{1}\psi_{B}c^{j}+t_{2}\psi_{B}c^{j-1},\\
\varepsilon c^{j}\psi_{B} & = & t_{1}\psi_{A}c^{j}+t_{2}\psi_{A}c^{j+1}+\alpha\psi_{B}c^{j-1}+\alpha\psi_{B}c^{j+1},
\end{eqnarray}
which can be rewritten as a matrix equation 
\begin{equation}
\varepsilon\begin{bmatrix}\psi_{A}\\
\psi_{B}
\end{bmatrix}=\begin{bmatrix}\alpha(c+1/c) & t_{1}+t_{2}/c\\
t_{1}+t_{2}c & \alpha(c+1/c)
\end{bmatrix}\begin{bmatrix}\psi_{A}\\
\psi_{B}
\end{bmatrix}.
\label{eq:cmatrix}
\end{equation}
The respective characteristic polynomial is \emph{quadratic} in $c$ (and quadratic in $\varepsilon$) if $\alpha=0$,
but becomes \emph{quartic} in $c$ (remaining quadratic in $\varepsilon$)  if $\alpha\neq0$. 
Therefore, for each energy value $\varepsilon$, we have two $c$ solutions 
[one corresponding to a left edge state, the other to a right edge state, 
see Fig.~\ref{fig:cvalues}(a)]  if $\alpha=0$, but four solutions if $\alpha\neq0$ [see Fig.~\ref{fig:cvalues}(b)]. 
The latter implies in certain energy intervals the existence
of two right (left) edge eigenstates in the infinite chain [states
with decay in the right (left) direction] that may be combined in
order to generate a single right (left) edge state with zero amplitudes
at all the sites of a unit cell of the chain, so that OBC
may be introduced at these sites \cite{Chen2020}. Note that while in the case
of the SSH model with OBC, a single condition selects the left edge
state from the set of left edge-like states of the infinite chain
and that condition is that of zero amplitude at a virtual site at
the left end (that will also impose zero amplitude in the respective
sublattice), when we have two left edge states in the infinite chain
with the same energy we need two conditions (these will be the zeros
of amplitude at the two virtual sites A and B to the left of the chain
so that the resulting eigenstate of the infinite chain does not feel
the OBC).

The eigenvalues $\varepsilon$ as functions of $c$ are
\begin{equation}
\varepsilon_{\pm}(c)=\frac{\alpha c^{2}+\alpha\pm\beta}{c},
\quad\beta=\sqrt{c (ct_{1}+t_{2}) (t_{1}+c t_{2}) }
\label{eq:energyedge}
\end{equation}
and the respective eigenstates are 
\begin{equation}
c^{j}\begin{bmatrix}\psi_{A} \\
\psi_{B}\end{bmatrix}
=c^j
\begin{bmatrix}1\\
\pm\dfrac{\beta}{ct_{1}+t_{2}}
\end{bmatrix}
=c^j
\begin{bmatrix}1\\
\pm\sqrt{c\gamma(c)}
\end{bmatrix},
\label{eq:camplitudesB}
\end{equation}
or equivalently
\begin{equation}
c^{j}\begin{bmatrix}\psi_{A} \\ \psi_{B}\end{bmatrix}
=c^j
\begin{bmatrix}
\pm\dfrac{ct_{1}+t_{2}}{\beta}
\\
1
\end{bmatrix}
=c^j
\begin{bmatrix}
\pm\dfrac{1}{\sqrt{c\gamma(c)}}\\
1
\end{bmatrix},
\label{eq:camplitudes}
\end{equation}
where 
\begin{equation}
\gamma(c) =\dfrac{t_1+ct_2}{ct_1+t_2}.
\end{equation} 
and where we have assumed that $c t_1+t_2$ is positive in the last equality of Eqs.~\ref{eq:camplitudes} and \ref{eq:camplitudesB}.
Note that the amplitudes do not depend on $\alpha$.

For $\alpha=0$, the edge state energy is determined imposing $\psi_A=0$ in Eq.~\ref{eq:camplitudes} or $\psi_B=0$ in Eq.~\ref{eq:camplitudesB}. This implies $c=-t_2/t_1$  or $c=-t_1/t_2$, and, depending on the value of the ratio $t_1/t_2$, these will be left or right edge states. In both situations, the respective energy given by Eq.~\ref{eq:energyedge} will be zero. These energies and the amplitude are indicated  in Figs.~\ref{fig:cvalues}(a) and (c), respectively, by the red dots.

\begin{figure}[t!]	\includegraphics[width=0.85\columnwidth]{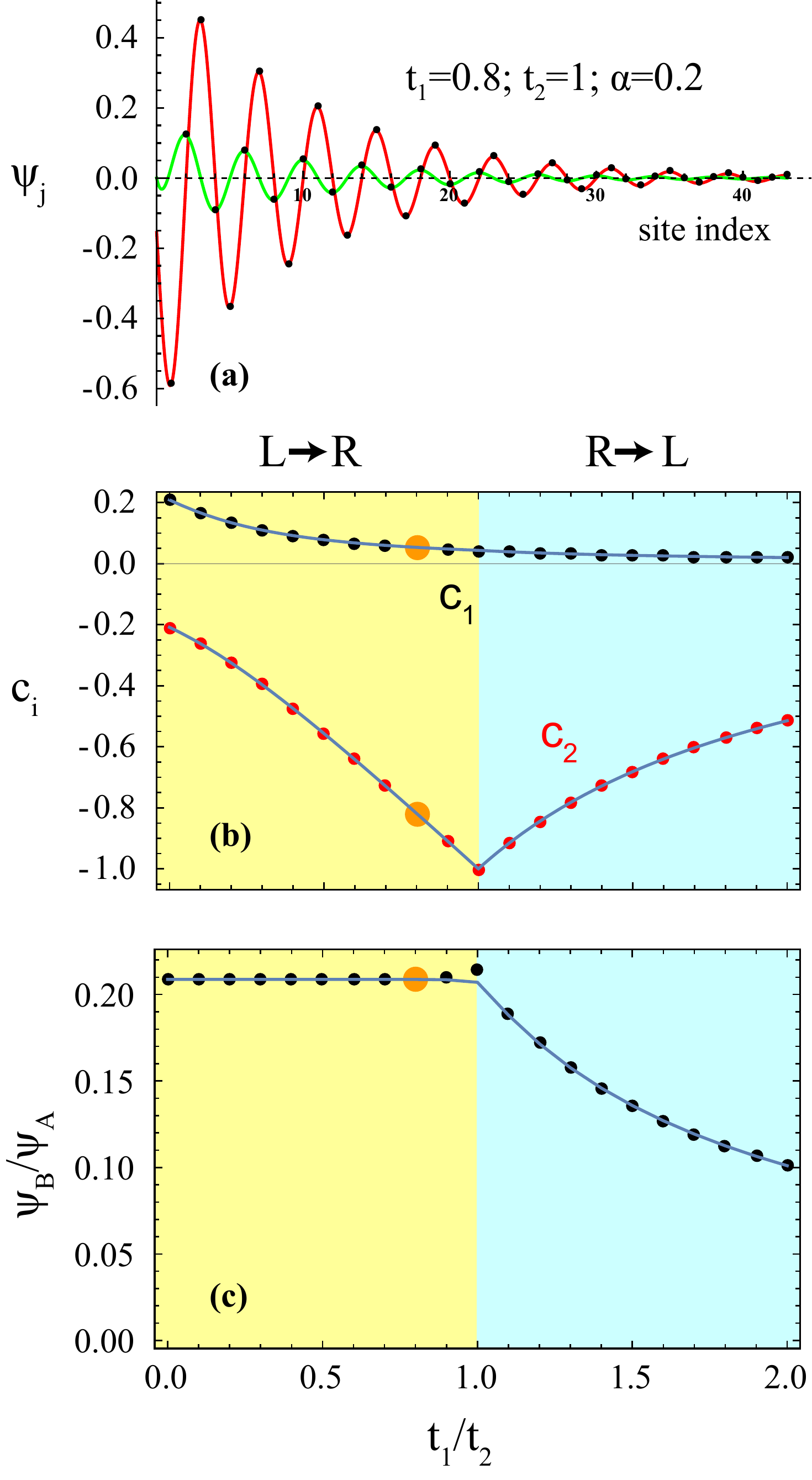}
	\caption{\label{fig:fits} (a) Edge state amplitudes for $t_1=0.8$, $t_2=1$, and $\alpha=0.2$ assuming the indexing of Fig.~\ref{fig:Usual-indexing-1}(a). The red and green curves are fits of the form $\psi_{A/B}(c_1^j-c_2^j)$ for respectively the A and B sublattices. (b) $c_1$ and $c_2$ values and (c)  $\psi_{B}/\psi_{A}$ extracted from the fits.  The dots in (b) and (c) are numerical data extracted from fits to the edge state amplitudes as shown in (a) and the lines are plotted from the theoretical results [Eq.~\ref{eq:cgammac} with $X(x)$ given by Eq.~\ref{eq:X} in the case of the $c$ values shown in (b) and Eq.~\ref{eq:camplitudesB} in the case of the amplitudes ratio shown in (c)].  For $t_1/t_2 >1$, one has a right edge state for chain endings as in Fig.~\ref{fig:Usual-indexing-1}(a) and in order to keep the $c$ values continuous, the indexing is flipped, that is, we index the sites from the right to the left. The edge state in (a) corresponds to the orange dots in (b) and (c). }
\end{figure}

For $\alpha\neq 0$, 
if the edge state has  two consecutive zeros of amplitude (this implies a unit
cell with zero amplitudes) in the chain, OBC may be introduced at the B site (the left edge) and one obtains
an eigenstate of the semi-infinite chain which is orthogonal to the
 bulk eigenstates. Let us determine the energy of such edge
state. As explained above, the existence of this edge state requires for
finite $\alpha$ the presence of four real $c$ solutions of the characteristic
equation, two of them corresponding to left edge-like states, see Fig.~\ref{fig:cvalues}(b) (one of the right edge solutions is outside the range of this plot). The left edge state is obtained combining two
left edge-like states with the same energy that  can be written as
\begin{eqnarray}
\ket{L_{1}}_{j} & = & c_{1}^{j}\begin{bmatrix}1\\
\psi_B(c_1)
\end{bmatrix},\\
\ket{L_{2}}_{j} & = & c_{2}^{j}\begin{bmatrix}1\\
\psi_B(c_2)
\end{bmatrix}.
\end{eqnarray}
Since we have NNN hoppings, in order for a linear combination of these
two states to be an eigenstate of the chain with OBC (or better, of a semi-infinite
chain), the amplitudes at both virtual sites (the unit cell with $j=0$)
must be zero, 
\begin{equation}
	a\begin{bmatrix}1\\
		\psi_B(c_1)
	\end{bmatrix}+b\begin{bmatrix}1\\
		\psi_B(c_2)
	\end{bmatrix}=\begin{bmatrix}0\\
		0
	\end{bmatrix}\Rightarrow\begin{cases}
		a=-b,\\
		\psi_B(c_1)=\psi_B(c_2),
	\end{cases}
\end{equation}
that is, $ \ket{edge} =\ket{L_{1}} -  \ket{L_{2}}$ and the energy of the edge state is determined by the condition
that at that energy one has two left (or two right) edge-like eigenstates
of the infinite chain that have the same unit cell amplitudes (apart
from the decaying factor). 
The relation between $c_1$ and $c_2$ is simple,
\begin{eqnarray}
\psi_B(c_1)=\psi_B(c_2) 
&\Rightarrow & c_1\gamma(c_1)=c_2\gamma(c_2) \nonumber  \\
&\Rightarrow  &
\left\{
\begin{matrix}
c_1=-\gamma(c_2) ,
\\
c_2=-\gamma(c_1) .
\end{matrix}
\right.
\end{eqnarray}
The last  two relations are equivalent.
This edge state can be written as 
\begin{eqnarray}
\ket{edge}_{j} & = & [c^{j}-(-\gamma(c))^{j}]\begin{bmatrix}1\\
\psi_B(c)
\end{bmatrix}.
\label{eq:edgestatefinal}
\end{eqnarray}
One can also conclude from Fig.~\ref{fig:cvalues}(d) that $c_{1}$ and
$c_{2}$ have opposite signs. 

In Fig.~\ref{fig:fits}(a), we show an example of amplitude profile of an edge state (for $t_1=0.8$, $t_2=1$, and $\alpha=0.2$) assuming the indexing of Fig.~\ref{fig:Usual-indexing-1}(a). The red and green curves are fits of the form $\psi_{A/B}(c_1^j-c_2^j)$ for respectively the A and B sublattices.  We considered just the largest amplitude  points for the fits since finite size effects should introduce deviations in the tail of the edge state (due to the present of the other boundary). In Fig.~\ref{fig:fits}(b), we show the $c_1$ and $c_2$ values and in Fig.~\ref{fig:fits}(c)  $\psi_{B}/\psi_{A}$ as $t_1/t_2$ grows for fixed $\alpha=0.2$.  The dots in Figs.~\ref{fig:fits}(b) and \ref{fig:fits}(c) are numerical data extracted from fits to the edge state amplitudes as shown in Fig.~\ref{fig:fits}(a) (the red and green fits generate the same values of $c_1$ and $c_2$) and the lines are plotted from the theoretical results, Eq.~\ref{eq:cgammac} with $X(x)$ given by Eq.~\ref{eq:X} in the case of the $c$ values and Eq.~\ref{eq:camplitudesB} in the case of the amplitudes ratio, as we explain below. An almost perfect fit is found with a small deviation at the topological transition point. For $t_1/t_2 >1$, one has a right edge state for chain endings as in Fig.~\ref{fig:Usual-indexing-1}(a) and in order to keep the $c$ values continuous, the indexing is flipped, that is, we index the sites from the right to the left. The edge state in Fig.~\ref{fig:fits}(a) corresponds to the orange dots in Figs.~\ref{fig:fits}(b) and \ref{fig:fits}(c).  

Note that Fig.~\ref{fig:fits}(c) shows that $\psi_B(c)$ (setting $\psi_A=1$) is a function of only  $\alpha/t_2$ for $t_1<t_2$, $\psi_B(c)=\pm X(\alpha/t_2)  $.  The apparent lack of symmetry for $t_1>t_2$ just reflects the fact that, in this region, $\psi_B(c)$ (setting $\psi_A=1$) is a function of $\alpha/t_1$ (due to the symmetry $t_1 \leftrightarrow t_2$ $\oplus$  L$\leftrightarrow$R), $\psi_B(c)=\pm\sqrt{c/\gamma(c)}=\pm  X(\alpha/t_1)$, and therefore $\psi_B(c)$ changes as $t_1$ grows for fixed $\alpha$. So we have 
\begin{equation}
\begin{split}
      \psi_B(c)
     & =\pm  
      \left\{
      \begin{matrix}
     \sqrt{c\gamma(c)}, \text{ for } t_1<t_2
      \\
      \sqrt{c/\gamma(c)}, \text{ for } t_1>t_2
      \end{matrix}
      \right.
      \\
     & =\pm  
      \left\{
      \begin{matrix}
      X(\alpha/t_2), \text{ for } t_1<t_2,
      \\
     X(\alpha/t_1), \text{ for } t_1>t_2.
      \end{matrix}
      \right.
      \end{split}
      \label{eq:cgammac}
\end{equation}
This equation determines the value of $c$ if the function $X(x)$ is known.

The function function $X(x)$  (and the value of $c$ in Eq.~\ref{eq:edgestatefinal}) is determined by the condition that the left edge states
$\ket{L_{1}}_{j} $ and $\ket{L_{2}}_{j} $ have the same energy, $\varepsilon_{\pm}(c_1)=\varepsilon_{\pm}(c_2) $ (and the same reasoning applies to right edge states) and this leads to
\begin{equation}
\varepsilon_{\pm}(c)=\varepsilon_{\pm}(-\gamma(c)).
\label{eq:energyequality}
\end{equation}
That is, the intersection of the curves $\varepsilon_{\pm}(c)$ and $\varepsilon_{\pm}(-\gamma(c))$ gives us the exact form of the edge state [see Fig.~\ref{fig:cvalues}(d)].
For $t_1/t_2=1/2$, one has $\varepsilon_{\pm}(c)=\alpha/t_2$.

Since $X(\alpha/t_2)$ is independent of $t_1$  for $t_1<t_2$, we may determine this function in the simple case $t_1=0$. For $t_1=0$, one has $\gamma(c)=c$, $\psi_B=\pm c$, $\beta=\vert c t_2\vert$ and $\varepsilon_{\pm}(c)=\varepsilon_{\pm}(-c) $ leads to
\begin{equation}
	c=X(\alpha/t_2),
\end{equation}
with
\begin{equation}
X(x)=\dfrac{\pm 1 \pm \sqrt{1-4x^2}}{2x},
\label{eq:X}
\end{equation}
where the four combinations must be considered (generating 2 left and 2 right edge states).
So, to conclude, with the above expression for $X(x)$, Eq.~\ref{eq:cgammac} determines the value of $c_1$ [$c_2$ being $-\gamma(c_1)$] and Eq.~\ref{eq:energyedge} gives the respective edge state energy.

\begin{figure}[t]
	\includegraphics[width=1\columnwidth]{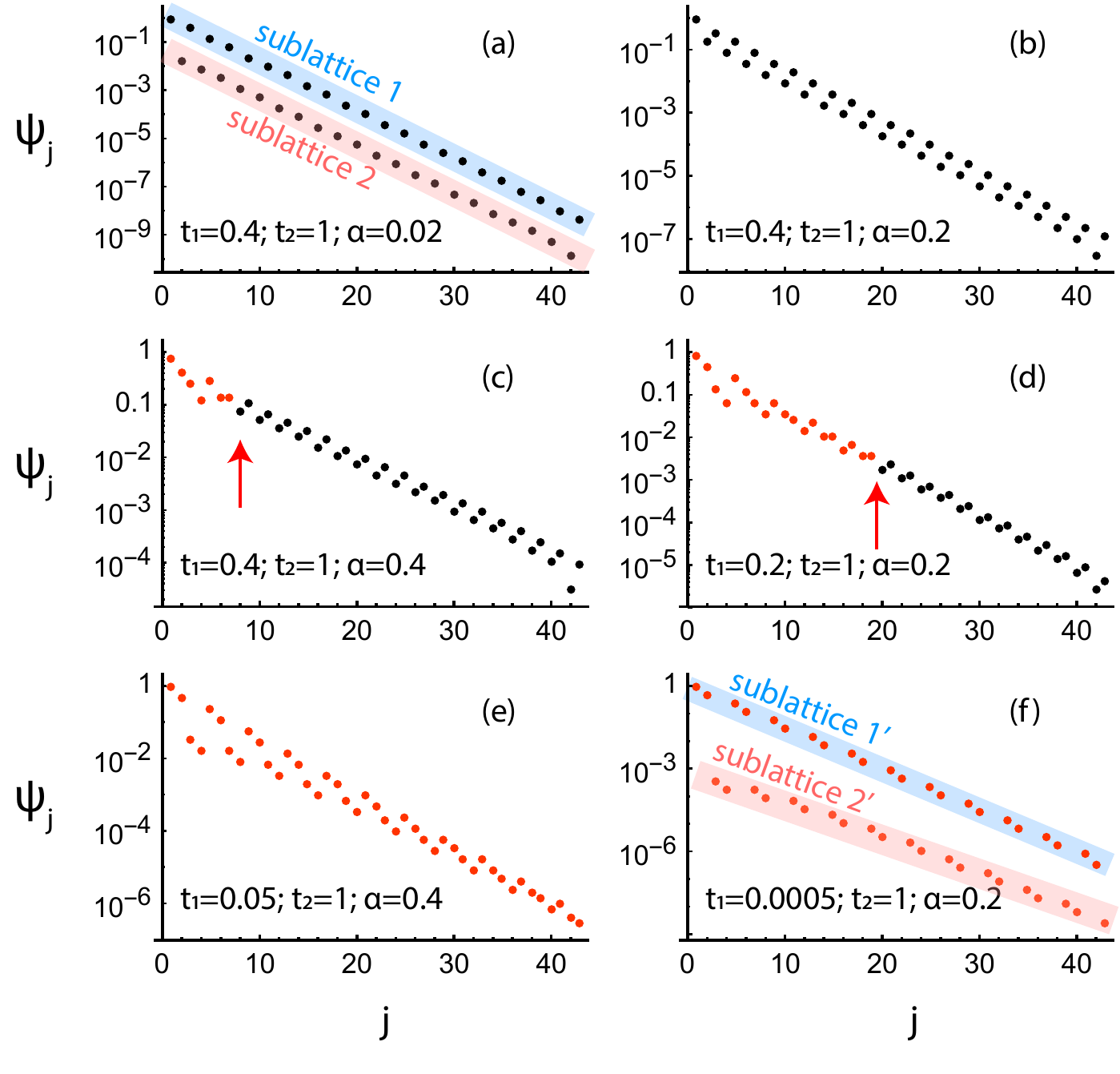}
	\caption{\label{fig:logplots} Log plot of the absolute values of the edge state amplitudes as a
		function of site index using the usual indexing, for several values
		of $t_{1}$and $\alpha$ ($t_{2}=1$). The change from the $t_1$-$t_2$ SSH chiral
		symmetry [reflected by the separation into two sublattices in (a)] to the other [corresponding to two different sublattices in (f)] is gradual, occurring  first at the edge of the edge state [see red arrows in (c) and (d)]. As $t_1$ approaches zero, $c_2\rightarrow -c_1$ and a doubling of the edge state ``unit cell'' is observed, reflecting the need for a different indexing.  If we replot these data using the new indexing of Fig.~\ref{fig:Usual-indexing-1}(c),  the behavior for small and large $t_1/\alpha$ is exchanged. In (a)-(f), the edge state amplitudes that show the doubling of the edge state “unit cell” (and reflect the existence of the sublattices 1’ and 2’) are colored in red. The edge state amplitudes given by the black dots reflect the usual bipartition of the SSH chain. }
\end{figure}

To complete our analysis of the effect of long range hopping on the topological edge states, we show in Fig.~\ref{fig:logplots} log plots of the absolute values of the edge state amplitudes as a
function of site index using the usual indexing, for several values
of $t_{1}$ and $\alpha$ ($t_{2}=1$). 
The  log-plot in Fig.~\ref{fig:logplots}(a), with $\alpha \ll  t_1$,  reflects the existence of the two sublattices associated with the indexing of Fig.~\ref{fig:Usual-indexing-1}(a) (due to the proximity to the SSH chiral symmetry point) while the log plot in Fig.~\ref{fig:logplots}(f) displays a doubling of the edge state ``unit cell'', reflecting the need for a different indexing with a new unit cell that generates the sublattices shown in Fig.~\ref{fig:logplots}(f).
The change from the ${\cal C}_1$ SSH chiral
symmetry [reflected by the separation into two sublattices in (a)] to the  ${\cal C}_2$ chiral
symmetry [corresponding to two different sublattices in (f)] is gradual, occurring  first at the edge of the edge state [see red arrows in (c) and (d)].   If we replot these data using the new indexing of Fig.~\ref{fig:Usual-indexing-1}(c),  the behavior for small and large $t_1/\alpha$ is exchanged.

\section{Conclusion}

The characterization of topological insulators depends strongly on its set of symmetries. In the case of translational invariant Hamiltonians, these symmetries can be written as a set of well known equations that involve the bulk Hamiltonian and the respective symmetry operator.
The bulk Hamiltonian reflects a choice of unit cell and it is usually assumed that any choice of unit cell will allow one to fully   characterize a topological insulator.
In this paper, we argue that this is not always correct and that  several  choices of unit cell or, equivalently, of site  indexing, should be considered in the case of linear chains with long range hoppings  in order to  describe their bulk topological behavior.
Otherwise, depending of the relative values of the hopping parameters, hidden symmetries may be present \cite{Li2015,Zurita2021,Fukui2013,Hou2016,Hou2017}.
We have exemplified the above arguments considering  the particular case of the extended SSH chain. In this model, the introduction of long-range hopping terms breaks in general the bipartite property causing the Hamiltonian to lose  its respective chiral symmetry, and a band inversion occurs in the band structure as the relative values of the hopping terms change.
We have shown this band inversion signals a crossover between hopping parameter regions of ``influence'' of different chiral symmetries and that depending on the choice of unit cell, these may be hidden  chiral symmetries. 

We have also shown  that, as a consequence of the long-range terms,  the edge states of the extended SSH chain become a linear combination of two edge-like states with different localization lengths but  equal amplitudes in the unit cell (apart from the decaying term). We have determined the exact form of these edge states and of the respective energy for any value of the next-nearest hopping term. 
When one of the nearest-neighbor hopping parameters becomes small compared with the next-nearest neighbor hopping parameter, the localization lengths become similar and this implies the gradual appearance of a chiral symmetry different from that of the SSH (edge states being protected by this new chiral symmetry when a nearest-neighbor hopping parameter is zero) that reflects a new sublattice distribution of the lattice sites  and the need for a new indexing of the lattice sites.
This need is also signaled by the inversion of one of the model bands. 

Similar behavior to that of the extended SSH chain should  be present in other bipartite 1D \cite{Marques2019,Kremer2020,Pelegri2019a,Pelegri2019b} and 2D \cite{Du2021,Liu2017,Obana2019,Madail2019} topological insulators if long-range hopping terms are introduced.
A simple  2D example where the approach of this paper can be applied is that of a plane of parallel extended SSH chains with uniform or staggered hopping terms between them.  
An interesting extension of these results should also be possible in the context of square-root topological insulators \cite{Arkinstall2017,Pelegri2019a, Kremer2020} or the recently proposed $2^n$-root topological insulators 
\cite{Marques2021a,Dias2021,Marques2021}.
We also expect that the introduction of long-range hopping terms in higher-order topological insulators \cite{Benalcazar2017,Schindler2018,Pelegri2019} to raise similar concerns about the possibility of hidden symmetries as well as the existence of multiple localization lengths in the corners states.
An open question is the effect of disorder in what concerns the linear combination of edge-like states with different localization lengths.

The experimental observation of the features described in this paper  should be  realizable  in artificial lattices such as ultracold atoms in optical lattices \cite{Leder2016}, photonic crystals \cite{Parappurath2020}, topolectrical circuits \cite{Imhof2018,Liu2019,Olekhno2020,Lee2018}, acoustic lattices \cite{Chen_2014,Zheng2019}, vacancy lattices  \cite{Drost2017}, or  atoms or molecules on surfaces \cite{Slot2017}. 
Since  the hopping terms are decreasing functions of the distance between sites (except in topolectrical circuits) the extended SSH realization should adopt the geometry of Fig.~\ref{fig:Usual-indexing-1}(a), but for example with a scalene  $t_1$-$t_2$-$\alpha$ triangle where the largest side corresponds to the smallest hopping parameter $t_{1,2}$.

\section*{Acknowledgments}\label{sec:acknowledments}
This work was developed within the scope of the Portuguese Institute for Nanostructures, Nanomodelling and Nanofabrication (i3N) projects UIDB/50025/2020 and UIDP/50025/2020. RGD and AMM acknowledge funding from FCT - Portuguese Foundation for Science and Technology through the project PTDC/FIS-MAC/29291/2017. AMM acknowledges financial support from the FCT through the work contract CDL-CTTRI-147-ARH/2018.

\bibliography{bibliografia2}

\begin{thebibliography}{53}%
\makeatletter
\providecommand \@ifxundefined [1]{%
 \@ifx{#1\undefined}
}%
\providecommand \@ifnum [1]{%
 \ifnum #1\expandafter \@firstoftwo
 \else \expandafter \@secondoftwo
 \fi
}%
\providecommand \@ifx [1]{%
 \ifx #1\expandafter \@firstoftwo
 \else \expandafter \@secondoftwo
 \fi
}%
\providecommand \natexlab [1]{#1}%
\providecommand \enquote  [1]{``#1''}%
\providecommand \bibnamefont  [1]{#1}%
\providecommand \bibfnamefont [1]{#1}%
\providecommand \citenamefont [1]{#1}%
\providecommand \href@noop [0]{\@secondoftwo}%
\providecommand \href [0]{\begingroup \@sanitize@url \@href}%
\providecommand \@href[1]{\@@startlink{#1}\@@href}%
\providecommand \@@href[1]{\endgroup#1\@@endlink}%
\providecommand \@sanitize@url [0]{\catcode `\\12\catcode `\$12\catcode
  `\&12\catcode `\#12\catcode `\^12\catcode `\_12\catcode `\%12\relax}%
\providecommand \@@startlink[1]{}%
\providecommand \@@endlink[0]{}%
\providecommand \url  [0]{\begingroup\@sanitize@url \@url }%
\providecommand \@url [1]{\endgroup\@href {#1}{\urlprefix }}%
\providecommand \urlprefix  [0]{URL }%
\providecommand \Eprint [0]{\href }%
\providecommand \doibase [0]{https://doi.org/}%
\providecommand \selectlanguage [0]{\@gobble}%
\providecommand \bibinfo  [0]{\@secondoftwo}%
\providecommand \bibfield  [0]{\@secondoftwo}%
\providecommand \translation [1]{[#1]}%
\providecommand \BibitemOpen [0]{}%
\providecommand \bibitemStop [0]{}%
\providecommand \bibitemNoStop [0]{.\EOS\space}%
\providecommand \EOS [0]{\spacefactor3000\relax}%
\providecommand \BibitemShut  [1]{\csname bibitem#1\endcsname}%
\let\auto@bib@innerbib\@empty
\bibitem [{\citenamefont {Altland}\ and\ \citenamefont
  {Zirnbauer}(1997)}]{Altland1997}%
  \BibitemOpen
  \bibfield  {author} {\bibinfo {author} {\bibfnamefont {A.}~\bibnamefont
  {Altland}}\ and\ \bibinfo {author} {\bibfnamefont {M.~R.}\ \bibnamefont
  {Zirnbauer}},\ }\bibfield  {title} {\bibinfo {title} {Nonstandard symmetry
  classes in mesoscopic normal-superconducting hybrid structures},\ }\href
  {https://doi.org/10.1103/PhysRevB.55.1142} {\bibfield  {journal} {\bibinfo
  {journal} {Phys. Rev. B}\ }\textbf {\bibinfo {volume} {55}},\ \bibinfo
  {pages} {1142} (\bibinfo {year} {1997})}\BibitemShut {NoStop}%
\bibitem [{\citenamefont {Chiu}\ \emph {et~al.}(2016)\citenamefont {Chiu},
  \citenamefont {Teo}, \citenamefont {Schnyder},\ and\ \citenamefont
  {Ryu}}]{Chiu2016}%
  \BibitemOpen
  \bibfield  {author} {\bibinfo {author} {\bibfnamefont {C.-K.}\ \bibnamefont
  {Chiu}}, \bibinfo {author} {\bibfnamefont {J.~C.}\ \bibnamefont {Teo}},
  \bibinfo {author} {\bibfnamefont {A.~P.}\ \bibnamefont {Schnyder}},\ and\
  \bibinfo {author} {\bibfnamefont {S.}~\bibnamefont {Ryu}},\ }\bibfield
  {title} {\bibinfo {title} {Classification of topological quantum matter with
  symmetries},\ }\bibfield  {journal} {\bibinfo  {journal} {Reviews of Modern
  Physics}\ }\textbf {\bibinfo {volume} {88}},\ \href
  {https://doi.org/10.1103/revmodphys.88.035005} {10.1103/revmodphys.88.035005}
  (\bibinfo {year} {2016})\BibitemShut {NoStop}%
\bibitem [{\citenamefont {Zak}(1989)}]{Zak1989}%
  \BibitemOpen
  \bibfield  {author} {\bibinfo {author} {\bibfnamefont {J.}~\bibnamefont
  {Zak}},\ }\bibfield  {title} {\bibinfo {title} {Berry's phase for energy
  bands in solids},\ }\href {https://doi.org/10.1103/physrevlett.62.2747}
  {\bibfield  {journal} {\bibinfo  {journal} {Physical Review Letters}\
  }\textbf {\bibinfo {volume} {62}},\ \bibinfo {pages} {2747} (\bibinfo {year}
  {1989})}\BibitemShut {NoStop}%
\bibitem [{\citenamefont {Asbóth}\ \emph {et~al.}(2016)\citenamefont
  {Asbóth}, \citenamefont {Oroszlány},\ and\ \citenamefont
  {Pályi}}]{Asboth2016}%
  \BibitemOpen
  \bibfield  {author} {\bibinfo {author} {\bibfnamefont {J.~K.}\ \bibnamefont
  {Asbóth}}, \bibinfo {author} {\bibfnamefont {L.}~\bibnamefont
  {Oroszlány}},\ and\ \bibinfo {author} {\bibfnamefont {A.}~\bibnamefont
  {Pályi}},\ }\bibfield  {title} {\bibinfo {title} {A short course on
  topological insulators},\ }\bibfield  {journal} {\bibinfo  {journal} {Lecture
  Notes in Physics}\ }\href {https://doi.org/10.1007/978-3-319-25607-8}
  {10.1007/978-3-319-25607-8} (\bibinfo {year} {2016})\BibitemShut {NoStop}%
\bibitem [{\citenamefont {Li}\ and\ \citenamefont {Chen}(2015)}]{Li2015}%
  \BibitemOpen
  \bibfield  {author} {\bibinfo {author} {\bibfnamefont {L.}~\bibnamefont
  {Li}}\ and\ \bibinfo {author} {\bibfnamefont {S.}~\bibnamefont {Chen}},\
  }\bibfield  {title} {\bibinfo {title} {Hidden-symmetry{\textendash}protected
  topological phases on a one-dimensional lattice},\ }\href
  {https://doi.org/10.1209/0295-5075/109/40006} {\bibfield  {journal} {\bibinfo
   {journal} {{EPL} (Europhysics Letters)}\ }\textbf {\bibinfo {volume}
  {109}},\ \bibinfo {pages} {40006} (\bibinfo {year} {2015})}\BibitemShut
  {NoStop}%
\bibitem [{\citenamefont {Zurita}\ \emph {et~al.}(2021)\citenamefont {Zurita},
  \citenamefont {Creffield},\ and\ \citenamefont {Platero}}]{Zurita2021}%
  \BibitemOpen
  \bibfield  {author} {\bibinfo {author} {\bibfnamefont {J.}~\bibnamefont
  {Zurita}}, \bibinfo {author} {\bibfnamefont {C.}~\bibnamefont {Creffield}},\
  and\ \bibinfo {author} {\bibfnamefont {G.}~\bibnamefont {Platero}},\
  }\bibfield  {title} {\bibinfo {title} {Tunable zero modes and quantum
  interferences in flat-band topological insulators},\ }\href
  {https://doi.org/10.22331/q-2021-11-25-591} {\bibfield  {journal} {\bibinfo
  {journal} {{Quantum}}\ }\textbf {\bibinfo {volume} {5}},\ \bibinfo {pages}
  {591} (\bibinfo {year} {2021})}\BibitemShut {NoStop}%
\bibitem [{\citenamefont {Fukui}\ \emph {et~al.}(2013)\citenamefont {Fukui},
  \citenamefont {Imura},\ and\ \citenamefont {Hatsugai}}]{Fukui2013}%
  \BibitemOpen
  \bibfield  {author} {\bibinfo {author} {\bibfnamefont {T.}~\bibnamefont
  {Fukui}}, \bibinfo {author} {\bibfnamefont {K.-I.}\ \bibnamefont {Imura}},\
  and\ \bibinfo {author} {\bibfnamefont {Y.}~\bibnamefont {Hatsugai}},\
  }\bibfield  {title} {\bibinfo {title} {Symmetry protected weak topological
  phases in a superlattice},\ }\href {https://doi.org/10.7566/JPSJ.82.073708}
  {\bibfield  {journal} {\bibinfo  {journal} {Journal of the Physical Society
  of Japan}\ }\textbf {\bibinfo {volume} {82}},\ \bibinfo {pages} {073708}
  (\bibinfo {year} {2013})},\ \Eprint
  {https://arxiv.org/abs/https://doi.org/10.7566/JPSJ.82.073708}
  {https://doi.org/10.7566/JPSJ.82.073708} \BibitemShut {NoStop}%
\bibitem [{\citenamefont {Hou}\ and\ \citenamefont {Chen}(2016)}]{Hou2016}%
  \BibitemOpen
  \bibfield  {author} {\bibinfo {author} {\bibfnamefont {J.-M.}\ \bibnamefont
  {Hou}}\ and\ \bibinfo {author} {\bibfnamefont {W.}~\bibnamefont {Chen}},\
  }\bibfield  {title} {\bibinfo {title} {Hidden-symmetry-protected quantum
  pseudo--spin hall effect in optical lattices},\ }\href
  {https://doi.org/10.1103/PhysRevA.93.063626} {\bibfield  {journal} {\bibinfo
  {journal} {Phys. Rev. A}\ }\textbf {\bibinfo {volume} {93}},\ \bibinfo
  {pages} {063626} (\bibinfo {year} {2016})}\BibitemShut {NoStop}%
\bibitem [{\citenamefont {Hou}\ and\ \citenamefont {Chen}(2017)}]{Hou2017}%
  \BibitemOpen
  \bibfield  {author} {\bibinfo {author} {\bibfnamefont {J.-M.}\ \bibnamefont
  {Hou}}\ and\ \bibinfo {author} {\bibfnamefont {W.}~\bibnamefont {Chen}},\
  }\bibfield  {title} {\bibinfo {title} {Hidden symmetry-protected ${Z}_{2}$
  topological insulator in a cubic lattice},\ }\href
  {https://doi.org/10.1103/PhysRevB.96.235108} {\bibfield  {journal} {\bibinfo
  {journal} {Phys. Rev. B}\ }\textbf {\bibinfo {volume} {96}},\ \bibinfo
  {pages} {235108} (\bibinfo {year} {2017})}\BibitemShut {NoStop}%
\bibitem [{\citenamefont {Hsu}\ and\ \citenamefont {Chen}(2020)}]{Hsu2020}%
  \BibitemOpen
  \bibfield  {author} {\bibinfo {author} {\bibfnamefont {H.-C.}\ \bibnamefont
  {Hsu}}\ and\ \bibinfo {author} {\bibfnamefont {T.-W.}\ \bibnamefont {Chen}},\
  }\bibfield  {title} {\bibinfo {title} {Topological anderson insulating phases
  in the long-range su-schrieffer-heeger model},\ }\href
  {https://doi.org/10.1103/PhysRevB.102.205425} {\bibfield  {journal} {\bibinfo
   {journal} {Phys. Rev. B}\ }\textbf {\bibinfo {volume} {102}},\ \bibinfo
  {pages} {205425} (\bibinfo {year} {2020})}\BibitemShut {NoStop}%
\bibitem [{\citenamefont {Maffei}\ \emph {et~al.}(2018)\citenamefont {Maffei},
  \citenamefont {Dauphin}, \citenamefont {Cardano}, \citenamefont
  {Lewenstein},\ and\ \citenamefont {Massignan}}]{Maffei2018}%
  \BibitemOpen
  \bibfield  {author} {\bibinfo {author} {\bibfnamefont {M.}~\bibnamefont
  {Maffei}}, \bibinfo {author} {\bibfnamefont {A.}~\bibnamefont {Dauphin}},
  \bibinfo {author} {\bibfnamefont {F.}~\bibnamefont {Cardano}}, \bibinfo
  {author} {\bibfnamefont {M.}~\bibnamefont {Lewenstein}},\ and\ \bibinfo
  {author} {\bibfnamefont {P.}~\bibnamefont {Massignan}},\ }\bibfield  {title}
  {\bibinfo {title} {Topological characterization of chiral models through
  their long time dynamics},\ }\href {https://doi.org/10.1088/1367-2630/aa9d4c}
  {\bibfield  {journal} {\bibinfo  {journal} {New Journal of Physics}\ }\textbf
  {\bibinfo {volume} {20}},\ \bibinfo {pages} {013023} (\bibinfo {year}
  {2018})}\BibitemShut {NoStop}%
\bibitem [{\citenamefont {Chen}\ and\ \citenamefont {Chiou}(2020)}]{Chen2020}%
  \BibitemOpen
  \bibfield  {author} {\bibinfo {author} {\bibfnamefont {B.-H.}\ \bibnamefont
  {Chen}}\ and\ \bibinfo {author} {\bibfnamefont {D.-W.}\ \bibnamefont
  {Chiou}},\ }\bibfield  {title} {\bibinfo {title} {An elementary rigorous
  proof of bulk-boundary correspondence in the generalized su-schrieffer-heeger
  model},\ }\href
  {https://doi.org/https://doi.org/10.1016/j.physleta.2019.126168} {\bibfield
  {journal} {\bibinfo  {journal} {Physics Letters A}\ }\textbf {\bibinfo
  {volume} {384}},\ \bibinfo {pages} {126168} (\bibinfo {year}
  {2020})}\BibitemShut {NoStop}%
\bibitem [{\citenamefont {Ahmadi}\ \emph {et~al.}(2020)\citenamefont {Ahmadi},
  \citenamefont {Abouie},\ and\ \citenamefont {Baeriswyl}}]{Ahmadi2020}%
  \BibitemOpen
  \bibfield  {author} {\bibinfo {author} {\bibfnamefont {N.}~\bibnamefont
  {Ahmadi}}, \bibinfo {author} {\bibfnamefont {J.}~\bibnamefont {Abouie}},\
  and\ \bibinfo {author} {\bibfnamefont {D.}~\bibnamefont {Baeriswyl}},\
  }\bibfield  {title} {\bibinfo {title} {Topological and nontopological
  features of generalized su-schrieffer-heeger models},\ }\href
  {https://doi.org/10.1103/PhysRevB.101.195117} {\bibfield  {journal} {\bibinfo
   {journal} {Phys. Rev. B}\ }\textbf {\bibinfo {volume} {101}},\ \bibinfo
  {pages} {195117} (\bibinfo {year} {2020})}\BibitemShut {NoStop}%
\bibitem [{\citenamefont {Zhang}\ and\ \citenamefont {Zhou}(2017)}]{Zhang2017}%
  \BibitemOpen
  \bibfield  {author} {\bibinfo {author} {\bibfnamefont {S.-L.}\ \bibnamefont
  {Zhang}}\ and\ \bibinfo {author} {\bibfnamefont {Q.}~\bibnamefont {Zhou}},\
  }\bibfield  {title} {\bibinfo {title} {Two-leg su-schrieffer-heeger chain
  with glide reflection symmetry},\ }\href
  {https://doi.org/10.1103/PhysRevA.95.061601} {\bibfield  {journal} {\bibinfo
  {journal} {Phys. Rev. A}\ }\textbf {\bibinfo {volume} {95}},\ \bibinfo
  {pages} {061601} (\bibinfo {year} {2017})}\BibitemShut {NoStop}%
\bibitem [{\citenamefont {Li}\ and\ \citenamefont
  {Miroshnichenko}(2019)}]{Li2019}%
  \BibitemOpen
  \bibfield  {author} {\bibinfo {author} {\bibfnamefont {C.}~\bibnamefont
  {Li}}\ and\ \bibinfo {author} {\bibfnamefont {A.~E.}\ \bibnamefont
  {Miroshnichenko}},\ }\bibfield  {title} {\bibinfo {title} {Extended ssh
  model: Non-local couplings and non-monotonous edge states},\ }\href
  {https://doi.org/10.3390/physics1010002} {\bibfield  {journal} {\bibinfo
  {journal} {Physics}\ }\textbf {\bibinfo {volume} {1}},\ \bibinfo {pages} {2}
  (\bibinfo {year} {2019})}\BibitemShut {NoStop}%
\bibitem [{\citenamefont {Xie}\ \emph {et~al.}(2019)\citenamefont {Xie},
  \citenamefont {Gou}, \citenamefont {Xiao}, \citenamefont {Gadway},\ and\
  \citenamefont {Yan}}]{Xie2019}%
  \BibitemOpen
  \bibfield  {author} {\bibinfo {author} {\bibfnamefont {D.}~\bibnamefont
  {Xie}}, \bibinfo {author} {\bibfnamefont {W.}~\bibnamefont {Gou}}, \bibinfo
  {author} {\bibfnamefont {T.}~\bibnamefont {Xiao}}, \bibinfo {author}
  {\bibfnamefont {B.}~\bibnamefont {Gadway}},\ and\ \bibinfo {author}
  {\bibfnamefont {B.}~\bibnamefont {Yan}},\ }\bibfield  {title} {\bibinfo
  {title} {Topological characterizations of an extended su-schrieffer-heeger
  model},\ }\href {https://doi.org/10.1038/s41534-019-0159-6} {\bibfield
  {journal} {\bibinfo  {journal} {npj Quantum Information}\ }\textbf {\bibinfo
  {volume} {5}},\ \bibinfo {pages} {55} (\bibinfo {year} {2019})}\BibitemShut
  {NoStop}%
\bibitem [{\citenamefont {Fu}\ \emph {et~al.}(2020)\citenamefont {Fu},
  \citenamefont {Fu}, \citenamefont {Zhang}, \citenamefont {Wang},
  \citenamefont {Zhao},\ and\ \citenamefont {Ke}}]{Fu2020}%
  \BibitemOpen
  \bibfield  {author} {\bibinfo {author} {\bibfnamefont {Z.}~\bibnamefont
  {Fu}}, \bibinfo {author} {\bibfnamefont {N.}~\bibnamefont {Fu}}, \bibinfo
  {author} {\bibfnamefont {H.}~\bibnamefont {Zhang}}, \bibinfo {author}
  {\bibfnamefont {Z.}~\bibnamefont {Wang}}, \bibinfo {author} {\bibfnamefont
  {D.}~\bibnamefont {Zhao}},\ and\ \bibinfo {author} {\bibfnamefont
  {S.}~\bibnamefont {Ke}},\ }\bibfield  {title} {\bibinfo {title} {Extended ssh
  model in non-hermitian waveguides with alternating real and imaginary
  couplings},\ }\bibfield  {journal} {\bibinfo  {journal} {Applied Sciences}\
  }\textbf {\bibinfo {volume} {10}},\ \href
  {https://doi.org/10.3390/app10103425} {10.3390/app10103425} (\bibinfo {year}
  {2020})\BibitemShut {NoStop}%
\bibitem [{\citenamefont {P\'erez-Gonz\'alez}\ \emph
  {et~al.}(2019)\citenamefont {P\'erez-Gonz\'alez}, \citenamefont {Bello},
  \citenamefont {G\'omez-Le\'on},\ and\ \citenamefont
  {Platero}}]{PerezGonzalez2019}%
  \BibitemOpen
  \bibfield  {author} {\bibinfo {author} {\bibfnamefont {B.}~\bibnamefont
  {P\'erez-Gonz\'alez}}, \bibinfo {author} {\bibfnamefont {M.}~\bibnamefont
  {Bello}}, \bibinfo {author} {\bibfnamefont {A.}~\bibnamefont
  {G\'omez-Le\'on}},\ and\ \bibinfo {author} {\bibfnamefont {G.}~\bibnamefont
  {Platero}},\ }\bibfield  {title} {\bibinfo {title} {Interplay between
  long-range hopping and disorder in topological systems},\ }\href
  {https://doi.org/10.1103/PhysRevB.99.035146} {\bibfield  {journal} {\bibinfo
  {journal} {Phys. Rev. B}\ }\textbf {\bibinfo {volume} {99}},\ \bibinfo
  {pages} {035146} (\bibinfo {year} {2019})}\BibitemShut {NoStop}%
\bibitem [{\citenamefont {Pérez-González}\ \emph {et~al.}(2018)\citenamefont
  {Pérez-González}, \citenamefont {Bello}, \citenamefont {Álvaro
  Gómez-León},\ and\ \citenamefont {Platero}}]{PerezGonzalez2018}%
  \BibitemOpen
  \bibfield  {author} {\bibinfo {author} {\bibfnamefont {B.}~\bibnamefont
  {Pérez-González}}, \bibinfo {author} {\bibfnamefont {M.}~\bibnamefont
  {Bello}}, \bibinfo {author} {\bibnamefont {Álvaro Gómez-León}},\ and\
  \bibinfo {author} {\bibfnamefont {G.}~\bibnamefont {Platero}},\ }\href@noop
  {} {\bibinfo {title} {Ssh model with long-range hoppings: topology, driving
  and disorder}} (\bibinfo {year} {2018}),\ \Eprint
  {https://arxiv.org/abs/1802.03973} {arXiv:1802.03973 [cond-mat.mes-hall]}
  \BibitemShut {NoStop}%
\bibitem [{\citenamefont {Longhi}(2018)}]{Longhi2018}%
  \BibitemOpen
  \bibfield  {author} {\bibinfo {author} {\bibfnamefont {S.}~\bibnamefont
  {Longhi}},\ }\bibfield  {title} {\bibinfo {title} {Probing one-dimensional
  topological phases in waveguide lattices with broken chiral symmetry},\
  }\href {https://doi.org/10.1364/OL.43.004639} {\bibfield  {journal} {\bibinfo
   {journal} {Opt. Lett.}\ }\textbf {\bibinfo {volume} {43}},\ \bibinfo {pages}
  {4639} (\bibinfo {year} {2018})}\BibitemShut {NoStop}%
\bibitem [{\citenamefont {Het\'enyi}\ \emph {et~al.}(2021)\citenamefont
  {Het\'enyi}, \citenamefont {Pulcu},\ and\ \citenamefont
  {Do\ifmmode~\breve{g}\else \u{g}\fi{}an}}]{Hetenyi2021}%
  \BibitemOpen
  \bibfield  {author} {\bibinfo {author} {\bibfnamefont {B.}~\bibnamefont
  {Het\'enyi}}, \bibinfo {author} {\bibfnamefont {Y.}~\bibnamefont {Pulcu}},\
  and\ \bibinfo {author} {\bibfnamefont {S.}~\bibnamefont
  {Do\ifmmode~\breve{g}\else \u{g}\fi{}an}},\ }\bibfield  {title} {\bibinfo
  {title} {Calculating the polarization in bipartite lattice models:
  Application to an extended su-schrieffer-heeger model},\ }\href
  {https://doi.org/10.1103/PhysRevB.103.075117} {\bibfield  {journal} {\bibinfo
   {journal} {Phys. Rev. B}\ }\textbf {\bibinfo {volume} {103}},\ \bibinfo
  {pages} {075117} (\bibinfo {year} {2021})}\BibitemShut {NoStop}%
\bibitem [{\citenamefont {Li}\ \emph {et~al.}(2014)\citenamefont {Li},
  \citenamefont {Xu},\ and\ \citenamefont {Chen}}]{Li2014}%
  \BibitemOpen
  \bibfield  {author} {\bibinfo {author} {\bibfnamefont {L.}~\bibnamefont
  {Li}}, \bibinfo {author} {\bibfnamefont {Z.}~\bibnamefont {Xu}},\ and\
  \bibinfo {author} {\bibfnamefont {S.}~\bibnamefont {Chen}},\ }\bibfield
  {title} {\bibinfo {title} {Topological phases of generalized
  su-schrieffer-heeger models},\ }\href
  {https://doi.org/10.1103/PhysRevB.89.085111} {\bibfield  {journal} {\bibinfo
  {journal} {Phys. Rev. B}\ }\textbf {\bibinfo {volume} {89}},\ \bibinfo
  {pages} {085111} (\bibinfo {year} {2014})}\BibitemShut {NoStop}%
\bibitem [{\citenamefont {Delplace}\ \emph {et~al.}(2011)\citenamefont
  {Delplace}, \citenamefont {Ullmo},\ and\ \citenamefont
  {Montambaux}}]{Delplace2011}%
  \BibitemOpen
  \bibfield  {author} {\bibinfo {author} {\bibfnamefont {P.}~\bibnamefont
  {Delplace}}, \bibinfo {author} {\bibfnamefont {D.}~\bibnamefont {Ullmo}},\
  and\ \bibinfo {author} {\bibfnamefont {G.}~\bibnamefont {Montambaux}},\
  }\bibfield  {title} {\bibinfo {title} {Zak phase and the existence of edge
  states in graphene},\ }\href {https://doi.org/10.1103/PhysRevB.84.195452}
  {\bibfield  {journal} {\bibinfo  {journal} {Phys. Rev. B}\ }\textbf {\bibinfo
  {volume} {84}},\ \bibinfo {pages} {195452} (\bibinfo {year}
  {2011})}\BibitemShut {NoStop}%
\bibitem [{\citenamefont {Banchi}\ and\ \citenamefont
  {Vaia}(2013)}]{Banchi2013}%
  \BibitemOpen
  \bibfield  {author} {\bibinfo {author} {\bibfnamefont {L.}~\bibnamefont
  {Banchi}}\ and\ \bibinfo {author} {\bibfnamefont {R.}~\bibnamefont {Vaia}},\
  }\bibfield  {title} {\bibinfo {title} {Spectral problem for quasi-uniform
  nearest-neighbor chains},\ }\href {https://doi.org/10.1063/1.4797477}
  {\bibfield  {journal} {\bibinfo  {journal} {Journal of Mathematical Physics}\
  }\textbf {\bibinfo {volume} {54}},\ \bibinfo {pages} {043501} (\bibinfo
  {year} {2013})}\BibitemShut {NoStop}%
\bibitem [{\citenamefont {H\"ugel}\ and\ \citenamefont
  {Paredes}(2014)}]{Huegel2014}%
  \BibitemOpen
  \bibfield  {author} {\bibinfo {author} {\bibfnamefont {D.}~\bibnamefont
  {H\"ugel}}\ and\ \bibinfo {author} {\bibfnamefont {B.}~\bibnamefont
  {Paredes}},\ }\bibfield  {title} {\bibinfo {title} {Chiral ladders and the
  edges of quantum hall insulators},\ }\href
  {https://doi.org/10.1103/PhysRevA.89.023619} {\bibfield  {journal} {\bibinfo
  {journal} {Phys. Rev. A}\ }\textbf {\bibinfo {volume} {89}},\ \bibinfo
  {pages} {023619} (\bibinfo {year} {2014})}\BibitemShut {NoStop}%
\bibitem [{\citenamefont {Duncan}\ \emph {et~al.}(2018)\citenamefont {Duncan},
  \citenamefont {\"Ohberg},\ and\ \citenamefont {Valiente}}]{Duncan2018}%
  \BibitemOpen
  \bibfield  {author} {\bibinfo {author} {\bibfnamefont {C.~W.}\ \bibnamefont
  {Duncan}}, \bibinfo {author} {\bibfnamefont {P.}~\bibnamefont {\"Ohberg}},\
  and\ \bibinfo {author} {\bibfnamefont {M.}~\bibnamefont {Valiente}},\
  }\bibfield  {title} {\bibinfo {title} {Exact edge, bulk, and bound states of
  finite topological systems},\ }\href
  {https://doi.org/10.1103/PhysRevB.97.195439} {\bibfield  {journal} {\bibinfo
  {journal} {Phys. Rev. B}\ }\textbf {\bibinfo {volume} {97}},\ \bibinfo
  {pages} {195439} (\bibinfo {year} {2018})}\BibitemShut {NoStop}%
\bibitem [{\citenamefont {Marques}\ and\ \citenamefont
  {Dias}(2019)}]{Marques2019}%
  \BibitemOpen
  \bibfield  {author} {\bibinfo {author} {\bibfnamefont {A.~M.}\ \bibnamefont
  {Marques}}\ and\ \bibinfo {author} {\bibfnamefont {R.~G.}\ \bibnamefont
  {Dias}},\ }\bibfield  {title} {\bibinfo {title} {One-dimensional topological
  insulators with noncentered inversion symmetry axis},\ }\href
  {https://doi.org/10.1103/PhysRevB.100.041104} {\bibfield  {journal} {\bibinfo
   {journal} {Phys. Rev. B}\ }\textbf {\bibinfo {volume} {100}},\ \bibinfo
  {pages} {041104(R)} (\bibinfo {year} {2019})}\BibitemShut {NoStop}%
\bibitem [{\citenamefont {Marques}\ and\ \citenamefont
  {Dias}(2020)}]{Marques2020}%
  \BibitemOpen
  \bibfield  {author} {\bibinfo {author} {\bibfnamefont {A.~M.}\ \bibnamefont
  {Marques}}\ and\ \bibinfo {author} {\bibfnamefont {R.~G.}\ \bibnamefont
  {Dias}},\ }\bibfield  {title} {\bibinfo {title} {Analytical solution of open
  crystalline linear 1d tight-binding models},\ }\href
  {https://doi.org/10.1088/1751-8121/ab6a6e} {\bibfield  {journal} {\bibinfo
  {journal} {Journal of Physics A: Mathematical and Theoretical}\ }\textbf
  {\bibinfo {volume} {53}},\ \bibinfo {pages} {075303} (\bibinfo {year}
  {2020})}\BibitemShut {NoStop}%
\bibitem [{\citenamefont {Fukui}(2020)}]{Fukui2020}%
  \BibitemOpen
  \bibfield  {author} {\bibinfo {author} {\bibfnamefont {T.}~\bibnamefont
  {Fukui}},\ }\bibfield  {title} {\bibinfo {title} {Theory of edge states based
  on the hermiticity of tight-binding hamiltonian operators},\ }\href
  {https://doi.org/10.1103/PhysRevResearch.2.043136} {\bibfield  {journal}
  {\bibinfo  {journal} {Phys. Rev. Research}\ }\textbf {\bibinfo {volume}
  {2}},\ \bibinfo {pages} {043136} (\bibinfo {year} {2020})}\BibitemShut
  {NoStop}%
\bibitem [{\citenamefont {Kremer}\ \emph {et~al.}(2020)\citenamefont {Kremer},
  \citenamefont {Petrides}, \citenamefont {Meyer}, \citenamefont {Heinrich},
  \citenamefont {Zilberberg},\ and\ \citenamefont {Szameit}}]{Kremer2020}%
  \BibitemOpen
  \bibfield  {author} {\bibinfo {author} {\bibfnamefont {M.}~\bibnamefont
  {Kremer}}, \bibinfo {author} {\bibfnamefont {I.}~\bibnamefont {Petrides}},
  \bibinfo {author} {\bibfnamefont {E.}~\bibnamefont {Meyer}}, \bibinfo
  {author} {\bibfnamefont {M.}~\bibnamefont {Heinrich}}, \bibinfo {author}
  {\bibfnamefont {O.}~\bibnamefont {Zilberberg}},\ and\ \bibinfo {author}
  {\bibfnamefont {A.}~\bibnamefont {Szameit}},\ }\bibfield  {title} {\bibinfo
  {title} {A square-root topological insulator with non-quantized indices
  realized with photonic aharonov-bohm cages},\ }\href
  {https://doi.org/10.1038/s41467-020-14692-4} {\bibfield  {journal} {\bibinfo
  {journal} {Nature Communications}\ }\textbf {\bibinfo {volume} {11}},\
  \bibinfo {pages} {907} (\bibinfo {year} {2020})}\BibitemShut {NoStop}%
\bibitem [{\citenamefont {Pelegr\'{\i}}\ \emph
  {et~al.}(2019{\natexlab{a}})\citenamefont {Pelegr\'{\i}}, \citenamefont
  {Marques}, \citenamefont {Dias}, \citenamefont {Daley}, \citenamefont
  {Ahufinger},\ and\ \citenamefont {Mompart}}]{Pelegri2019a}%
  \BibitemOpen
  \bibfield  {author} {\bibinfo {author} {\bibfnamefont {G.}~\bibnamefont
  {Pelegr\'{\i}}}, \bibinfo {author} {\bibfnamefont {A.~M.}\ \bibnamefont
  {Marques}}, \bibinfo {author} {\bibfnamefont {R.~G.}\ \bibnamefont {Dias}},
  \bibinfo {author} {\bibfnamefont {A.~J.}\ \bibnamefont {Daley}}, \bibinfo
  {author} {\bibfnamefont {V.}~\bibnamefont {Ahufinger}},\ and\ \bibinfo
  {author} {\bibfnamefont {J.}~\bibnamefont {Mompart}},\ }\bibfield  {title}
  {\bibinfo {title} {Topological edge states with ultracold atoms carrying
  orbital angular momentum in a diamond chain},\ }\href
  {https://doi.org/10.1103/PhysRevA.99.023612} {\bibfield  {journal} {\bibinfo
  {journal} {Phys. Rev. A}\ }\textbf {\bibinfo {volume} {99}},\ \bibinfo
  {pages} {023612} (\bibinfo {year} {2019}{\natexlab{a}})}\BibitemShut
  {NoStop}%
\bibitem [{\citenamefont {Pelegr\'{\i}}\ \emph
  {et~al.}(2019{\natexlab{b}})\citenamefont {Pelegr\'{\i}}, \citenamefont
  {Marques}, \citenamefont {Dias}, \citenamefont {Daley}, \citenamefont
  {Mompart},\ and\ \citenamefont {Ahufinger}}]{Pelegri2019b}%
  \BibitemOpen
  \bibfield  {author} {\bibinfo {author} {\bibfnamefont {G.}~\bibnamefont
  {Pelegr\'{\i}}}, \bibinfo {author} {\bibfnamefont {A.~M.}\ \bibnamefont
  {Marques}}, \bibinfo {author} {\bibfnamefont {R.~G.}\ \bibnamefont {Dias}},
  \bibinfo {author} {\bibfnamefont {A.~J.}\ \bibnamefont {Daley}}, \bibinfo
  {author} {\bibfnamefont {J.}~\bibnamefont {Mompart}},\ and\ \bibinfo {author}
  {\bibfnamefont {V.}~\bibnamefont {Ahufinger}},\ }\bibfield  {title} {\bibinfo
  {title} {Topological edge states and aharanov-bohm caging with ultracold
  atoms carrying orbital angular momentum},\ }\href
  {https://doi.org/10.1103/PhysRevA.99.023613} {\bibfield  {journal} {\bibinfo
  {journal} {Phys. Rev. A}\ }\textbf {\bibinfo {volume} {99}},\ \bibinfo
  {pages} {023613} (\bibinfo {year} {2019}{\natexlab{b}})}\BibitemShut
  {NoStop}%
\bibitem [{\citenamefont {Du}\ \emph {et~al.}(2021)\citenamefont {Du},
  \citenamefont {Li}, \citenamefont {Lu},\ and\ \citenamefont
  {Zhang}}]{Du2021}%
  \BibitemOpen
  \bibfield  {author} {\bibinfo {author} {\bibfnamefont {T.}~\bibnamefont
  {Du}}, \bibinfo {author} {\bibfnamefont {Y.}~\bibnamefont {Li}}, \bibinfo
  {author} {\bibfnamefont {H.}~\bibnamefont {Lu}},\ and\ \bibinfo {author}
  {\bibfnamefont {H.}~\bibnamefont {Zhang}},\ }\bibfield  {title} {\bibinfo
  {title} {Effects of correlations on phase diagrams of the two-dimensional
  su–schrieffer–heeger model with the larger topological invariant},\
  }\href {https://doi.org/https://doi.org/10.1016/j.physe.2021.114884}
  {\bibfield  {journal} {\bibinfo  {journal} {Physica E: Low-dimensional
  Systems and Nanostructures}\ }\textbf {\bibinfo {volume} {134}},\ \bibinfo
  {pages} {114884} (\bibinfo {year} {2021})}\BibitemShut {NoStop}%
\bibitem [{\citenamefont {Liu}\ and\ \citenamefont
  {Wakabayashi}(2017)}]{Liu2017}%
  \BibitemOpen
  \bibfield  {author} {\bibinfo {author} {\bibfnamefont {F.}~\bibnamefont
  {Liu}}\ and\ \bibinfo {author} {\bibfnamefont {K.}~\bibnamefont
  {Wakabayashi}},\ }\bibfield  {title} {\bibinfo {title} {Novel topological
  phase with a zero berry curvature},\ }\href
  {https://doi.org/10.1103/PhysRevLett.118.076803} {\bibfield  {journal}
  {\bibinfo  {journal} {Phys. Rev. Lett.}\ }\textbf {\bibinfo {volume} {118}},\
  \bibinfo {pages} {076803} (\bibinfo {year} {2017})}\BibitemShut {NoStop}%
\bibitem [{\citenamefont {Obana}\ \emph {et~al.}(2019)\citenamefont {Obana},
  \citenamefont {Liu},\ and\ \citenamefont {Wakabayashi}}]{Obana2019}%
  \BibitemOpen
  \bibfield  {author} {\bibinfo {author} {\bibfnamefont {D.}~\bibnamefont
  {Obana}}, \bibinfo {author} {\bibfnamefont {F.}~\bibnamefont {Liu}},\ and\
  \bibinfo {author} {\bibfnamefont {K.}~\bibnamefont {Wakabayashi}},\
  }\bibfield  {title} {\bibinfo {title} {Topological edge states in the
  su-schrieffer-heeger model},\ }\href
  {https://doi.org/10.1103/PhysRevB.100.075437} {\bibfield  {journal} {\bibinfo
   {journal} {Phys. Rev. B}\ }\textbf {\bibinfo {volume} {100}},\ \bibinfo
  {pages} {075437} (\bibinfo {year} {2019})}\BibitemShut {NoStop}%
\bibitem [{\citenamefont {Madail}\ \emph {et~al.}(2019)\citenamefont {Madail},
  \citenamefont {Flannigan}, \citenamefont {Marques}, \citenamefont {Daley},\
  and\ \citenamefont {Dias}}]{Madail2019}%
  \BibitemOpen
  \bibfield  {author} {\bibinfo {author} {\bibfnamefont {L.}~\bibnamefont
  {Madail}}, \bibinfo {author} {\bibfnamefont {S.}~\bibnamefont {Flannigan}},
  \bibinfo {author} {\bibfnamefont {A.~M.}\ \bibnamefont {Marques}}, \bibinfo
  {author} {\bibfnamefont {A.~J.}\ \bibnamefont {Daley}},\ and\ \bibinfo
  {author} {\bibfnamefont {R.~G.}\ \bibnamefont {Dias}},\ }\bibfield  {title}
  {\bibinfo {title} {Enhanced localization and protection of topological edge
  states due to geometric frustration},\ }\href
  {https://doi.org/10.1103/PhysRevB.100.125123} {\bibfield  {journal} {\bibinfo
   {journal} {Phys. Rev. B}\ }\textbf {\bibinfo {volume} {100}},\ \bibinfo
  {pages} {125123} (\bibinfo {year} {2019})}\BibitemShut {NoStop}%
\bibitem [{\citenamefont {Arkinstall}\ \emph {et~al.}(2017)\citenamefont
  {Arkinstall}, \citenamefont {Teimourpour}, \citenamefont {Feng},
  \citenamefont {El-Ganainy},\ and\ \citenamefont
  {Schomerus}}]{Arkinstall2017}%
  \BibitemOpen
  \bibfield  {author} {\bibinfo {author} {\bibfnamefont {J.}~\bibnamefont
  {Arkinstall}}, \bibinfo {author} {\bibfnamefont {M.~H.}\ \bibnamefont
  {Teimourpour}}, \bibinfo {author} {\bibfnamefont {L.}~\bibnamefont {Feng}},
  \bibinfo {author} {\bibfnamefont {R.}~\bibnamefont {El-Ganainy}},\ and\
  \bibinfo {author} {\bibfnamefont {H.}~\bibnamefont {Schomerus}},\ }\bibfield
  {title} {\bibinfo {title} {Topological tight-binding models from nontrivial
  square roots},\ }\href {https://doi.org/10.1103/PhysRevB.95.165109}
  {\bibfield  {journal} {\bibinfo  {journal} {Phys. Rev. B}\ }\textbf {\bibinfo
  {volume} {95}},\ \bibinfo {pages} {165109} (\bibinfo {year}
  {2017})}\BibitemShut {NoStop}%
\bibitem [{\citenamefont {Marques}\ \emph {et~al.}(2021)\citenamefont
  {Marques}, \citenamefont {Madail},\ and\ \citenamefont
  {Dias}}]{Marques2021a}%
  \BibitemOpen
  \bibfield  {author} {\bibinfo {author} {\bibfnamefont {A.~M.}\ \bibnamefont
  {Marques}}, \bibinfo {author} {\bibfnamefont {L.}~\bibnamefont {Madail}},\
  and\ \bibinfo {author} {\bibfnamefont {R.~G.}\ \bibnamefont {Dias}},\
  }\bibfield  {title} {\bibinfo {title} {One-dimensional ${2}^{n}$-root
  topological insulators and superconductors},\ }\href
  {https://doi.org/10.1103/PhysRevB.103.235425} {\bibfield  {journal} {\bibinfo
   {journal} {Phys. Rev. B}\ }\textbf {\bibinfo {volume} {103}},\ \bibinfo
  {pages} {235425} (\bibinfo {year} {2021})}\BibitemShut {NoStop}%
\bibitem [{\citenamefont {Dias}\ and\ \citenamefont
  {Marques}(2021)}]{Dias2021}%
  \BibitemOpen
  \bibfield  {author} {\bibinfo {author} {\bibfnamefont {R.~G.}\ \bibnamefont
  {Dias}}\ and\ \bibinfo {author} {\bibfnamefont {A.~M.}\ \bibnamefont
  {Marques}},\ }\bibfield  {title} {\bibinfo {title} {Matryoshka approach to
  sine-cosine topological models},\ }\href
  {https://doi.org/10.1103/PhysRevB.103.245112} {\bibfield  {journal} {\bibinfo
   {journal} {Phys. Rev. B}\ }\textbf {\bibinfo {volume} {103}},\ \bibinfo
  {pages} {245112} (\bibinfo {year} {2021})}\BibitemShut {NoStop}%
\bibitem [{\citenamefont {Marques}\ and\ \citenamefont
  {Dias}(2021)}]{Marques2021}%
  \BibitemOpen
  \bibfield  {author} {\bibinfo {author} {\bibfnamefont {A.~M.}\ \bibnamefont
  {Marques}}\ and\ \bibinfo {author} {\bibfnamefont {R.~G.}\ \bibnamefont
  {Dias}},\ }\href@noop {} {\bibinfo {title} {$2^n$-root weak, chern and
  higher-order topological insulators and $2^n$-root topological semimetals}}
  (\bibinfo {year} {2021}),\ \Eprint {https://arxiv.org/abs/2107.13974}
  {arXiv:2107.13974 [cond-mat.mes-hall]} \BibitemShut {NoStop}%
\bibitem [{\citenamefont {Benalcazar}\ \emph {et~al.}(2017)\citenamefont
  {Benalcazar}, \citenamefont {Bernevig},\ and\ \citenamefont
  {Hughes}}]{Benalcazar2017}%
  \BibitemOpen
  \bibfield  {author} {\bibinfo {author} {\bibfnamefont {W.~A.}\ \bibnamefont
  {Benalcazar}}, \bibinfo {author} {\bibfnamefont {B.~A.}\ \bibnamefont
  {Bernevig}},\ and\ \bibinfo {author} {\bibfnamefont {T.~L.}\ \bibnamefont
  {Hughes}},\ }\bibfield  {title} {\bibinfo {title} {Electric multipole
  moments, topological multipole moment pumping, and chiral hinge states in
  crystalline insulators},\ }\href {https://doi.org/10.1103/PhysRevB.96.245115}
  {\bibfield  {journal} {\bibinfo  {journal} {Phys. Rev. B}\ }\textbf {\bibinfo
  {volume} {96}},\ \bibinfo {pages} {245115} (\bibinfo {year}
  {2017})}\BibitemShut {NoStop}%
\bibitem [{\citenamefont {Schindler}\ \emph {et~al.}(2018)\citenamefont
  {Schindler}, \citenamefont {Cook}, \citenamefont {Vergniory}, \citenamefont
  {Wang}, \citenamefont {Parkin}, \citenamefont {Bernevig},\ and\ \citenamefont
  {Neupert}}]{Schindler2018}%
  \BibitemOpen
  \bibfield  {author} {\bibinfo {author} {\bibfnamefont {F.}~\bibnamefont
  {Schindler}}, \bibinfo {author} {\bibfnamefont {A.~M.}\ \bibnamefont {Cook}},
  \bibinfo {author} {\bibfnamefont {M.~G.}\ \bibnamefont {Vergniory}}, \bibinfo
  {author} {\bibfnamefont {Z.}~\bibnamefont {Wang}}, \bibinfo {author}
  {\bibfnamefont {S.~S.~P.}\ \bibnamefont {Parkin}}, \bibinfo {author}
  {\bibfnamefont {B.~A.}\ \bibnamefont {Bernevig}},\ and\ \bibinfo {author}
  {\bibfnamefont {T.}~\bibnamefont {Neupert}},\ }\bibfield  {title} {\bibinfo
  {title} {Higher-order topological insulators},\ }\bibfield  {journal}
  {\bibinfo  {journal} {Science Advances}\ }\textbf {\bibinfo {volume} {4}},\
  \href {https://doi.org/10.1126/sciadv.aat0346} {10.1126/sciadv.aat0346}
  (\bibinfo {year} {2018})\BibitemShut {NoStop}%
\bibitem [{\citenamefont {Pelegr\'{\i}}\ \emph
  {et~al.}(2019{\natexlab{c}})\citenamefont {Pelegr\'{\i}}, \citenamefont
  {Marques}, \citenamefont {Ahufinger}, \citenamefont {Mompart},\ and\
  \citenamefont {Dias}}]{Pelegri2019}%
  \BibitemOpen
  \bibfield  {author} {\bibinfo {author} {\bibfnamefont {G.}~\bibnamefont
  {Pelegr\'{\i}}}, \bibinfo {author} {\bibfnamefont {A.~M.}\ \bibnamefont
  {Marques}}, \bibinfo {author} {\bibfnamefont {V.}~\bibnamefont {Ahufinger}},
  \bibinfo {author} {\bibfnamefont {J.}~\bibnamefont {Mompart}},\ and\ \bibinfo
  {author} {\bibfnamefont {R.~G.}\ \bibnamefont {Dias}},\ }\bibfield  {title}
  {\bibinfo {title} {Second-order topological corner states with ultracold
  atoms carrying orbital angular momentum in optical lattices},\ }\href
  {https://doi.org/10.1103/PhysRevB.100.205109} {\bibfield  {journal} {\bibinfo
   {journal} {Phys. Rev. B}\ }\textbf {\bibinfo {volume} {100}},\ \bibinfo
  {pages} {205109} (\bibinfo {year} {2019}{\natexlab{c}})}\BibitemShut
  {NoStop}%
\bibitem [{\citenamefont {Leder}\ \emph {et~al.}(2016)\citenamefont {Leder},
  \citenamefont {Grossert}, \citenamefont {Sitta}, \citenamefont {Genske},
  \citenamefont {Rosch},\ and\ \citenamefont {Weitz}}]{Leder2016}%
  \BibitemOpen
  \bibfield  {author} {\bibinfo {author} {\bibfnamefont {M.}~\bibnamefont
  {Leder}}, \bibinfo {author} {\bibfnamefont {C.}~\bibnamefont {Grossert}},
  \bibinfo {author} {\bibfnamefont {L.}~\bibnamefont {Sitta}}, \bibinfo
  {author} {\bibfnamefont {M.}~\bibnamefont {Genske}}, \bibinfo {author}
  {\bibfnamefont {A.}~\bibnamefont {Rosch}},\ and\ \bibinfo {author}
  {\bibfnamefont {M.}~\bibnamefont {Weitz}},\ }\bibfield  {title} {\bibinfo
  {title} {Real-space imaging of a topologically protected edge state with
  ultracold atoms in an amplitude-chirped optical lattice},\ }\href
  {https://doi.org/10.1038/ncomms13112} {\bibfield  {journal} {\bibinfo
  {journal} {Nature Communications}\ }\textbf {\bibinfo {volume} {7}},\
  \bibinfo {pages} {13112} (\bibinfo {year} {2016})}\BibitemShut {NoStop}%
\bibitem [{\citenamefont {Parappurath}\ \emph {et~al.}(2020)\citenamefont
  {Parappurath}, \citenamefont {Alpeggiani}, \citenamefont {Kuipers},\ and\
  \citenamefont {Verhagen}}]{Parappurath2020}%
  \BibitemOpen
  \bibfield  {author} {\bibinfo {author} {\bibfnamefont {N.}~\bibnamefont
  {Parappurath}}, \bibinfo {author} {\bibfnamefont {F.}~\bibnamefont
  {Alpeggiani}}, \bibinfo {author} {\bibfnamefont {L.}~\bibnamefont
  {Kuipers}},\ and\ \bibinfo {author} {\bibfnamefont {E.}~\bibnamefont
  {Verhagen}},\ }\bibfield  {title} {\bibinfo {title} {Direct observation of
  topological edge states in silicon photonic crystals: Spin, dispersion, and
  chiral routing},\ }\href {https://doi.org/10.1126/sciadv.aaw4137} {\bibfield
  {journal} {\bibinfo  {journal} {Science Advances}\ }\textbf {\bibinfo
  {volume} {6}},\ \bibinfo {pages} {eaaw4137} (\bibinfo {year}
  {2020})}\BibitemShut {NoStop}%
\bibitem [{\citenamefont {Imhof}\ \emph {et~al.}(2018)\citenamefont {Imhof}
  \emph {et~al.}}]{Imhof2018}%
  \BibitemOpen
  \bibfield  {author} {\bibinfo {author} {\bibfnamefont {S.}~\bibnamefont
  {Imhof}} \emph {et~al.},\ }\bibfield  {title} {\bibinfo {title}
  {Topolectrical-circuit realization of topological corner modes},\ }\href
  {https://doi.org/10.1038/s41567-018-0246-1} {\bibfield  {journal} {\bibinfo
  {journal} {Nature Physics}\ }\textbf {\bibinfo {volume} {14}},\ \bibinfo
  {pages} {925} (\bibinfo {year} {2018})}\BibitemShut {NoStop}%
\bibitem [{\citenamefont {Liu}\ \emph {et~al.}(2019)\citenamefont {Liu},
  \citenamefont {Gao}, \citenamefont {Zhang}, \citenamefont {Ma}, \citenamefont
  {Zhang}, \citenamefont {Liu}, \citenamefont {Xiang}, \citenamefont {Cui},\
  and\ \citenamefont {Zhang}}]{Liu2019}%
  \BibitemOpen
  \bibfield  {author} {\bibinfo {author} {\bibfnamefont {S.}~\bibnamefont
  {Liu}}, \bibinfo {author} {\bibfnamefont {W.}~\bibnamefont {Gao}}, \bibinfo
  {author} {\bibfnamefont {Q.}~\bibnamefont {Zhang}}, \bibinfo {author}
  {\bibfnamefont {S.}~\bibnamefont {Ma}}, \bibinfo {author} {\bibfnamefont
  {L.}~\bibnamefont {Zhang}}, \bibinfo {author} {\bibfnamefont
  {C.}~\bibnamefont {Liu}}, \bibinfo {author} {\bibfnamefont {Y.~J.}\
  \bibnamefont {Xiang}}, \bibinfo {author} {\bibfnamefont {T.~J.}\ \bibnamefont
  {Cui}},\ and\ \bibinfo {author} {\bibfnamefont {S.}~\bibnamefont {Zhang}},\
  }\bibfield  {title} {\bibinfo {title} {Topologically protected edge state in
  two-dimensional su-schrieffer-heeger circuit},\ }\href
  {https://doi.org/10.34133/2019/8609875} {\bibfield  {journal} {\bibinfo
  {journal} {Research}\ }\textbf {\bibinfo {volume} {2019}},\ \bibinfo {pages}
  {8609875} (\bibinfo {year} {2019})}\BibitemShut {NoStop}%
\bibitem [{\citenamefont {Olekhno}\ \emph {et~al.}(2020)\citenamefont
  {Olekhno}, \citenamefont {Kretov}, \citenamefont {Stepanenko}, \citenamefont
  {Ivanova}, \citenamefont {Yaroshenko}, \citenamefont {Puhtina}, \citenamefont
  {Filonov}, \citenamefont {Cappello}, \citenamefont {Matekovits},\ and\
  \citenamefont {Gorlach}}]{Olekhno2020}%
  \BibitemOpen
  \bibfield  {author} {\bibinfo {author} {\bibfnamefont {N.~A.}\ \bibnamefont
  {Olekhno}}, \bibinfo {author} {\bibfnamefont {E.~I.}\ \bibnamefont {Kretov}},
  \bibinfo {author} {\bibfnamefont {A.~A.}\ \bibnamefont {Stepanenko}},
  \bibinfo {author} {\bibfnamefont {P.~A.}\ \bibnamefont {Ivanova}}, \bibinfo
  {author} {\bibfnamefont {V.~V.}\ \bibnamefont {Yaroshenko}}, \bibinfo
  {author} {\bibfnamefont {E.~M.}\ \bibnamefont {Puhtina}}, \bibinfo {author}
  {\bibfnamefont {D.~S.}\ \bibnamefont {Filonov}}, \bibinfo {author}
  {\bibfnamefont {B.}~\bibnamefont {Cappello}}, \bibinfo {author}
  {\bibfnamefont {L.}~\bibnamefont {Matekovits}},\ and\ \bibinfo {author}
  {\bibfnamefont {M.~A.}\ \bibnamefont {Gorlach}},\ }\bibfield  {title}
  {\bibinfo {title} {Topological edge states of interacting photon pairs
  emulated in a topolectrical circuit},\ }\href
  {https://doi.org/10.1038/s41467-020-14994-7} {\bibfield  {journal} {\bibinfo
  {journal} {Nature Communications}\ }\textbf {\bibinfo {volume} {11}},\
  \bibinfo {pages} {1436} (\bibinfo {year} {2020})}\BibitemShut {NoStop}%
\bibitem [{\citenamefont {Lee}\ \emph {et~al.}(2018)\citenamefont {Lee},
  \citenamefont {Imhof}, \citenamefont {Berger}, \citenamefont {Bayer},
  \citenamefont {Brehm}, \citenamefont {Molenkamp}, \citenamefont {Kiessling},\
  and\ \citenamefont {Thomale}}]{Lee2018}%
  \BibitemOpen
  \bibfield  {author} {\bibinfo {author} {\bibfnamefont {C.~H.}\ \bibnamefont
  {Lee}}, \bibinfo {author} {\bibfnamefont {S.}~\bibnamefont {Imhof}}, \bibinfo
  {author} {\bibfnamefont {C.}~\bibnamefont {Berger}}, \bibinfo {author}
  {\bibfnamefont {F.}~\bibnamefont {Bayer}}, \bibinfo {author} {\bibfnamefont
  {J.}~\bibnamefont {Brehm}}, \bibinfo {author} {\bibfnamefont {L.~W.}\
  \bibnamefont {Molenkamp}}, \bibinfo {author} {\bibfnamefont {T.}~\bibnamefont
  {Kiessling}},\ and\ \bibinfo {author} {\bibfnamefont {R.}~\bibnamefont
  {Thomale}},\ }\bibfield  {title} {\bibinfo {title} {Topolectrical circuits},\
  }\href {https://doi.org/10.1038/s42005-018-0035-2} {\bibfield  {journal}
  {\bibinfo  {journal} {Communications Physics}\ }\textbf {\bibinfo {volume}
  {1}},\ \bibinfo {pages} {39} (\bibinfo {year} {2018})}\BibitemShut {NoStop}%
\bibitem [{\citenamefont {Chen}\ \emph {et~al.}(2014)\citenamefont {Chen},
  \citenamefont {Upadhyaya},\ and\ \citenamefont {Vitelli}}]{Chen_2014}%
  \BibitemOpen
  \bibfield  {author} {\bibinfo {author} {\bibfnamefont {B.~G.-g.}\
  \bibnamefont {Chen}}, \bibinfo {author} {\bibfnamefont {N.}~\bibnamefont
  {Upadhyaya}},\ and\ \bibinfo {author} {\bibfnamefont {V.}~\bibnamefont
  {Vitelli}},\ }\bibfield  {title} {\bibinfo {title} {Nonlinear conduction via
  solitons in a topological mechanical insulator},\ }\href
  {https://doi.org/10.1073/pnas.1405969111} {\bibfield  {journal} {\bibinfo
  {journal} {Proceedings of the National Academy of Sciences}\ }\textbf
  {\bibinfo {volume} {111}},\ \bibinfo {pages} {13004–13009} (\bibinfo {year}
  {2014})}\BibitemShut {NoStop}%
\bibitem [{\citenamefont {Zheng}\ \emph {et~al.}(2019)\citenamefont {Zheng},
  \citenamefont {Achilleos}, \citenamefont {Richoux}, \citenamefont
  {Theocharis},\ and\ \citenamefont {Pagneux}}]{Zheng2019}%
  \BibitemOpen
  \bibfield  {author} {\bibinfo {author} {\bibfnamefont {L.-Y.}\ \bibnamefont
  {Zheng}}, \bibinfo {author} {\bibfnamefont {V.}~\bibnamefont {Achilleos}},
  \bibinfo {author} {\bibfnamefont {O.}~\bibnamefont {Richoux}}, \bibinfo
  {author} {\bibfnamefont {G.}~\bibnamefont {Theocharis}},\ and\ \bibinfo
  {author} {\bibfnamefont {V.}~\bibnamefont {Pagneux}},\ }\bibfield  {title}
  {\bibinfo {title} {Observation of edge waves in a two-dimensional
  su-schrieffer-heeger acoustic network},\ }\href
  {https://doi.org/10.1103/PhysRevApplied.12.034014} {\bibfield  {journal}
  {\bibinfo  {journal} {Phys. Rev. Applied}\ }\textbf {\bibinfo {volume}
  {12}},\ \bibinfo {pages} {034014} (\bibinfo {year} {2019})}\BibitemShut
  {NoStop}%
\bibitem [{\citenamefont {Drost}\ \emph {et~al.}(2017)\citenamefont {Drost},
  \citenamefont {Ojanen}, \citenamefont {Harju},\ and\ \citenamefont
  {Liljeroth}}]{Drost2017}%
  \BibitemOpen
  \bibfield  {author} {\bibinfo {author} {\bibfnamefont {R.}~\bibnamefont
  {Drost}}, \bibinfo {author} {\bibfnamefont {T.}~\bibnamefont {Ojanen}},
  \bibinfo {author} {\bibfnamefont {A.}~\bibnamefont {Harju}},\ and\ \bibinfo
  {author} {\bibfnamefont {P.}~\bibnamefont {Liljeroth}},\ }\bibfield  {title}
  {\bibinfo {title} {Topological states in engineered atomic lattices},\ }\href
  {https://doi.org/10.1038/nphys4080} {\bibfield  {journal} {\bibinfo
  {journal} {Nature Physics}\ }\textbf {\bibinfo {volume} {13}},\ \bibinfo
  {pages} {668} (\bibinfo {year} {2017})}\BibitemShut {NoStop}%
\bibitem [{\citenamefont {Slot}\ \emph {et~al.}(2017)\citenamefont {Slot},
  \citenamefont {Gardenier}, \citenamefont {Jacobse}, \citenamefont {van
  Miert}, \citenamefont {Kempkes}, \citenamefont {Zevenhuizen}, \citenamefont
  {Smith}, \citenamefont {Vanmaekelbergh},\ and\ \citenamefont
  {Swart}}]{Slot2017}%
  \BibitemOpen
  \bibfield  {author} {\bibinfo {author} {\bibfnamefont {M.~R.}\ \bibnamefont
  {Slot}}, \bibinfo {author} {\bibfnamefont {T.~S.}\ \bibnamefont {Gardenier}},
  \bibinfo {author} {\bibfnamefont {P.~H.}\ \bibnamefont {Jacobse}}, \bibinfo
  {author} {\bibfnamefont {G.~C.~P.}\ \bibnamefont {van Miert}}, \bibinfo
  {author} {\bibfnamefont {S.~N.}\ \bibnamefont {Kempkes}}, \bibinfo {author}
  {\bibfnamefont {S.~J.~M.}\ \bibnamefont {Zevenhuizen}}, \bibinfo {author}
  {\bibfnamefont {C.~M.}\ \bibnamefont {Smith}}, \bibinfo {author}
  {\bibfnamefont {D.}~\bibnamefont {Vanmaekelbergh}},\ and\ \bibinfo {author}
  {\bibfnamefont {I.}~\bibnamefont {Swart}},\ }\bibfield  {title} {\bibinfo
  {title} {Experimental realization and characterization of an electronic lieb
  lattice},\ }\href {https://doi.org/10.1038/nphys4105} {\bibfield  {journal}
  {\bibinfo  {journal} {Nature Physics}\ }\textbf {\bibinfo {volume} {13}},\
  \bibinfo {pages} {672} (\bibinfo {year} {2017})}\BibitemShut {NoStop}%
\end{thebibliography}%

\end{document}